\documentclass[unsortedaddress,onecolumn, nofootinbib,10pt]{revtex4} 
\usepackage[a4paper,bindingoffset=0.2in, left=0.5in,right=0.8in,top=1in,bottom=1in, footskip=.25in]{geometry}
\usepackage[utf8]{inputenc}
\usepackage{amsmath,empheq,color}
\usepackage{amsfonts}
\usepackage{amsthm}
\usepackage{amssymb}
\usepackage{graphicx} 
\usepackage{hyperref}
\usepackage[normalem]{ulem}
\usepackage{epigraph}
\usepackage{mathrsfs}
\usepackage{bbm}
\usepackage{wrapfig}
\usepackage{lineno}
\usepackage{scalerel,stackengine}
\usepackage[T1]{fontenc}
\usepackage{epigraph}
\usepackage{cancel}
\usepackage{soul}
\usepackage{ulem}
\usepackage[textwidth=1.7cm,textsize=small]{tod onotes}

	\newcommand{\ncd}{\newcommand}
	\newcommand{\vsp}{\vspace{0.4cm}}

	\ncd{\mrm}    {\mathrm}
	\ncd{\beq} {\begin{equation}}
	\ncd{\eeq} {\end{equation}}

	\def\d{{\rm d}}
	\def\D{{\rm D}}
	\def\i{{\rm i}}

	\def\C{{\mathbb{C}}}
	\def\R{{\mathbb{R}}}

	\def\h{{\mathcal{H}}}
	
	\def\S{{\mathcal{S}}}
	
	\def\a{{\mathcal{A}}}
	\def\Ri{{\mathcal{C}}}

	\def\sl{{\mathfrak{sl}}}

	\def\su{{\mathfrak{su}}}
	
	\def\sp{{\mathfrak{sp}}}

	\def\D{{\mathfrak{D}}}
	\def\T{{\mathfrak{T}}}

	\def\e{{\rm e}}
	
	\def\Tr{{\rm Tr}}

\begin{document}

\title{GKLS Vector Field Dynamics for Gaussian States}

\author{Hans Cruz-Prado}
\email{hans@ciencias.unam.mx}
\affiliation{Departamento de F\'isica, Facultad de Ciencias, Universidad Nacional Aut\'onoma de M\'exico, A. P. 70543, Ciudad de M\'exico, 04510 M\'exico.}

\author{Octavio Casta\~nos}
\email{On the sabbatical leave at the Universidad de Granada, Espa\~na; ocasta@nucleares.unam.mx} 
\affiliation{Instituto de Ciencias Nucleares, Universidad Nacional Aut\'onoma de M\'exico, A. P. 70543, Ciudad de M\'exico 04510, M\'exico.}

\author{Giuseppe Marmo}%
\email{marmo@na.infn.it}
\affiliation{ 
INFN-Sezione di Napoli, Complesso Universitario di Monte S. Angelo Edificio 6, via Cintia, 80126 Napoli, Italy.}
\affiliation{ 
Dipartimento di Fisica ``E. Pancini'', Universit\`a di Napoli Federico II, Complesso Universitario di Monte S. Angelo Edificio 6, via Cintia, 80126 Napoli, Italy.}

\author{Francisco Nettel}
\email{fnettel@ciencias.unam.mx} 
\affiliation{
Departamento de F\'isica, Facultad de Ciencias, Universidad Nacional Aut\'onoma de M\'exico, A. P. 70543, Ciudad de M\'exico, 04510 M\'exico.}


\begin{abstract} {\bf{Abstrasct}}:
We construct the vector field associated to the GKLS generator for systems described by Gaussian states .
This vector field is defined on the dual space of the algebra of operators, restricted to operators quadratic in position and momentum. 
It is shown that the GKLS dynamics accepts a \emph{decomposition principle}, that is, this vector field can be decomposed in three parts, a conservative Hamiltonian component, a gradient-like, and a Choi--Kraus or jump vector field. The two last terms are considered a  ``perturbation'' associated with dissipation.
Examples are presented for a harmonic oscillator with different dissipation terms.
\end{abstract}

\maketitle


\section{Motivation and previous works}

The geometric formulation of quantum mechanics makes use not only of structures and methods used in classical mechanics (symplectic structures and Poisson brackets) but also of Jordan brackets and complex structures, suggesting a powerful framework for approaching both, the conceptual and the mathematical foundations of quantum mechanics.

Such a geometric approach consists in describing all the fundamental properties of the quantum theory through the use of \emph{geometric structures} on an appropriate Hilbert manifold, that is, the Born  probabilistic interpretation requires to work on manifolds rather than on vector spaces; moreover, by analogy with methods in information theory, the analog of ``statistical models'' in the space of probability distributions is played here by generalized coherent states, i.e., submanifolds of quantum states \cite{Strocchi-1966, Cantoni-1978, Kramer-1980, Cirelli-1983}.
In the Hilbert-space approach to quantum mechanics, Born probabilistic interpretation requires dealing with normalized vectors, losing therefore, the linear structure and requiring the use of Hilbert manifolds instead of Hilbert spaces.
Thus, in this manifold point of view, the Hermitian inner product should be replaced with a corresponding Hermitian tensor field.
Then, if we consider the manifold as a real differential manifold, already at the level of the Hilbert space, the Hermitian tensor splits into a symplectic structure, a symmetric metric tensor, and a connecting complex structure.
On the other hand, the contravariant version of these tensor fields, provides us with bidifferential operators on the Hilbert manifold~\cite{Grabowski-2005, Aniello-2011}.
These bidifferential operators, when applied to quadratic functions (expectation-value functions of Hermitian operators) define a skew-symmetric bracket (Poisson bracket) which reproduces the commutator of operators appearing in the quadratic form, and a symmetric bracket that corresponds to the Jordan product of Hermitian operators.
In this manner, we dispose of the operators and consider only functions (quadratic).
In the space of pure states, which are normalized vectors up to a phase, i.e., in the so-called \emph{complex-projective space}, we do not have a notion of quadratic function because there is not a linear structure. 
However, in this space, we have a symplectic structure and an associated Poisson bracket, along with a symmetric tensor and an associated bidifferential operator.
With the Poisson bracket, we define the Hamiltonian vector field and select those functions whose Hamiltonian vector fields are also Killing vector fields for the symmetric tensor (the Fubini-Study metric).
It may be shown that these functions are exactly those that correspond to the quadratic functions on the initial Hilbert space.
In this manner we recover the $C^\ast$-algebra of bounded operators but now realized as an algebra of special functions on the space of pure states (complex projective space).

Finally, by using the momentum map associated with those symplectic transformations which  also preserve the Jordan brackets, we map the complex projective space into the minimal orbit of the coadjoint action of the unitary group on the dual vector space of its Lie algebra.
In this dual vector space (considered as the space of Hermitian operators) we identify the space of all quantum states as non-negative, normalized, Hermitian operators.
Again, the realization in terms of operators allows us to introduce a Poisson bracket and a Jordan bracket on the dual space of linear functions.
The pullback of these functions using the momentum map gives the quadratic functions on the Hilbert space or the special functions, i.e, K\"ahlerian functions according to the symplectic action on the Hilbert space or the symplectic action on the complex projective space) \cite{Chru?ci?sk-2009}.

Nowadays, there is a growing interest in the geometric description of open quantum systems and their dynamics; specifically, there have been great advances in the geometric study of the general master equation governing the Markovian dynamics of finite quantum systems \cite{Grabowski-2005, Aniello-2011, Ciaglia-2017, Chruscinski-2019}. 
From here onward, we restrict our considerations to finite dimensional Hilbert spaces.
Following \cite{Ciaglia-2017}, we can establish the kinematics for the $n$-levels quantum systems. 
In this case, we are dealing with systems whose observables are elements of a finite $C^\ast$-algebra, i.e., being $\mathcal{M}_n(\C)$ the $C^\ast$-algebra of $(n \times n)$ complex matrices, then the observables are elements of the finite algebra $\a_n = \mathcal{M}_n(\C)$, for $n \geq 2$. 

Thus, the space of observables $\D_n$ is identified with the subset of $\a_n$ consisting of \emph{self-adjoint} elements, i.e. $\D_n \subset \a_n$. In addition, $\a_n$ possess a natural Hilbert inner product $\langle \hat{a} \, , \, \hat{b} \rangle := \Tr\{ \hat{a}^\dagger \hat{b} \}$ and hence $\a_n$ is a complex Hilbert space. If $\a^\ast_n$ is the dual space of $\a_n$, it is known that for every $\xi \in \a^\ast_n$ there is a unique $\hat{\xi} \in \a_n$ such that $\xi(\hat{a}) = \langle \hat{\xi} \, , \, \hat{a} \rangle$ for all $\hat{a} \in \a_n$; then, the space of states $\mathcal{S}$ of $\a_n$ is defined as
	\beq
	\S:= \{ \rho \in \D_n^\ast \subset \a_n^\ast \, | \, \rho(\hat{a} \hat{a}^\dagger) \geq 0 , \rho(\hat{\mathbb{I}}) = 1\}  \, ,
	\eeq 
where $\mathbb{I} \in \a_n$ is the identity operator and ${\hat a} \in \D_n$. 
Therefore, for each quantum state $\xi \equiv \rho \in \a_n^\ast$ there is a corresponding $\hat{\rho} \in \a_n$ defined as the self-adjoint semi-positive operator such that $\Tr\{ \hat{\rho} \, \hat{\mathbb{I}} \} = 1$, meaning that $\hat{\rho}$ is a density operator. 
Furthermore, from the one-to-one correspondence between elements of $\D_n^\ast$ and elements of $\D_n$, it follows that $\S$ may be decomposed as
	\beq
	\S =  \bigsqcup_{r=1}^n \S_r \, ,
	\eeq 
with
	\beq
	\S_r =  \{ \rho \in \S \, | \, \text{rank}(\rho) = r \} \, ,
	\eeq
where $\text{rank}(\rho)$ denotes the rank of $\rho$ defined as the matrix rank of the density matrix $\hat{\rho} \in \D_n$. Then, as has been shown in Refs. \cite{Ciaglia-2017, Chruscinski-2019}, $\S_r$ is a homegeneous space defined by a non-linear action of the Lie group
	\beq
	SL(\a_n) := \{ g \in \a_n | \det(g) = 1\} \, ,
	\eeq	
and therefore is a differential manifold.
The group $SL(\a_n)$ is a complexification of the special unitary group
	\beq
	SU(\a_n) := \{ \mathbf{U} \in \a_n | \mathbf{U}\mathbf{U}^\dagger = \mathbb{I},  \det(\mathbf{U}) = 1\} \, ,
	\eeq
where $SU(\a_n) \subset SL(\a_n)$, such that, on each $\S_k$ there is an action of this compact Lie subgroup 
and the orbits of this action are known in quantum theory as \emph{isospectral manifolds}.
In particular, the manifold of pure states, i.e. $\text{rank}(\rho) = 1$, turns out to be a homogeneous space for $SU(\a_n)$ and $SL(\a_n)$. 

Once the manifold of states is characterized, relevant geometric structures can be introduced. Each isospectral manifold is endowed with a K\"ahler structure, thus, through the symplectic form, it is possible to define a Hamiltonian dynamics on these manifolds which corresponds to the unitary evolution, see Refs. \cite{Grabowski-2005, Ciaglia-2017}. 
In this work, we will introduce more general geometric structures on $\S$, so a more general dynamical evolution can be defined than the symplectic one. 
This new dynamics is defined for the entire manifold of states, hence it may describe also the evolution of open quantum systems.
For instance, the dynamics associated with the Markovian evolution introduced in Refs.~\cite{Gorini-1976, Lindblad-1976} by means of the Gorini--Kossakowski--Lindblad--Sudarshan (GKLS) master equation 
	\beq \label{GKLS-generator}
	L(\hat{\rho}) =  - \frac{\i}{\hbar} [\hat{H}, \hat{\rho}]_{-} 
	- \frac{1}{2} \sum^3_j [ \hat{v}^\dagger_j \, \hat{v}_j , \hat{\rho} ]_{+}
	+ \sum^3_j \hat{v}^\dagger_j \, \hat{\rho} \, \hat{v}_j \, ,
	\eeq
where $\hat{\rho}$ is the density operator associated to a quantum state $\rho \in \S$, $\hat{H} \in \D_n$ is the Hamiltonian operator and the family of operators $ \hat{v}_j \in \a_n$ allows to describe dissipation by means of the second term and the third term; here, $[ \, \cdot \, , \, \cdot \, ]_{-}$ and $[ \, \cdot \, , \, \cdot \, ]_{+}$ denote the commutator and the anti-commutator in $\a_n$.
It can be seen that the operator $L$ is trace preserving and it is also possible to show that it is a completely positive map. 
The geometric description of the GKLS master equation means to describe the GKLS equation of motion through a vector field $\Gamma$ in the affine space $\T_n^1 \subset \D^\ast_n $ defined as
	\beq
	\T_n^1 = \{ \xi \in \D_n^\ast \, | \, \Tr\{ \hat{\xi} \, \mathbb{I} \} = 1 \} \, .
	\eeq 
Using the Lie--Jordan structure on $\D_n$ it is possible to introduce a ``gradient'' vector field such that the infinitesimal generator $\Gamma$ can decomposed as
	\beq \label{Gamma-n-level}
	\Gamma = X_H + Y_b + Z_\mathcal{K} \, ,
	\eeq  
where $X_H$ is the Hamiltonian vector field associated with the first term of the GKLS generator in \eqref{GKLS-generator} and $Y_b + Z_\mathcal{K}$ are related to the dissipative part of the  GKLS generator given by the second and third terms in \eqref{GKLS-generator}.
Then, the Hamiltonian vector fields are tangent to $\T_n^1$, more precisely, they are tangent to the manifolds of quantum states. 
On the other hand, the vector field $Y_b + Z_\mathcal{K}$ generates a dynamical evolution that \emph{changes} the rank and the spectrum of the density matrix, that is, it represents a dissipation term in the dynamics.

There are two types of systems in what refers to the encoding of the quantum information: i) systems with a discrete spectrum possibly countable, i.e. in the form of q-bits or q-dits and ii) systems with an a continuous spectrum such as those given in the form of the position or momentum representations.
For instance, one may consider Gaussian states that emerge naturally in Hamiltonians quadratic in the position and momentum variables~\cite{Malkin-1969, Malkin-1973}, or those associated with Hamiltonians which are described employing algebraic structures whose states are called generalized coherent states~\cite{Perelomov-1972, Arecchi-1972, Onofri-1975}.
All of them, even the non-linear coherent states~\cite{Manko-1997, Aniello-2000, Aniello-2009, Cruz-2021}, constitute examples of quantum states whose properties may be described by finite-dimensional smooth manifolds. 

In particular, the generalized coherent states solutions to the Schr\"odinger equation for Hamiltonians quadratic in the position and momentum operators, have been extensively studied due to their broad application in quantum optics and quantum technologies. In these cases, one may express the density operator in the position representation in the general form by means of a Gaussian kernel representation (a two-point function)
	\beq \label{Gaus-Den-Mat}
	\langle q' | \, \hat{\rho} \, | q \rangle 
	=  \frac{1}{\sqrt{2\, \pi \, \sigma^2_{q}}}  
	\exp \Bigg\{
	 \frac{1}{2 \, \sigma^2_{q} }
	\left[
	\frac{\sigma_{qp}}{\hbar} (q - q' ) - \frac{\i}{2} \, ( q + q' - 2 \langle \hat{q} \rangle )
	\right]^2 - \frac{ \sigma^2_{p} }{ 2 \, \hbar^2} (q - q')^2  + \frac{\i}{\hbar} \, \langle \hat{p} \rangle \, (q - q') 				\Bigg\} \, .
	\eeq
These normalized positive functional are uniquely parametrized by the first $(\langle \hat{q} \rangle , \langle \hat{p} \rangle)$ and second  $( \sigma_{q}^2,  \sigma_{p}^2,  \sigma_{qp} )$ moments. The second moments are constrained by the saturated Robertson--Schr\"odinger uncertainty relation, i.e., $\sigma_{q}^2\sigma_{p}^2 - \sigma^2_{qp} = \frac{\hbar}{4}$.  
The first and second moments parametrize the space of states defining a finite-dimensional manifold with a quantum evolution that can be described through a symplectic evolution, for details see Ref.~\cite{Cruz-2021}. 

For the Gaussian kernel given in \eqref{Gaus-Den-Mat} is possible to construct the GKLS dynamical vector field $\Gamma$ defined in the space of parameters, which may be decomposed as the vector field in \eqref{Gamma-n-level}, Hamiltonian, gradient and jump vector field. 
To see this, notice that the density matrix in \eqref{Gaus-Den-Mat} can describe non-pure states, as has been remarked in Ref.~\cite{Ferraro-2005}, where the degree of purity is given by a parameter $r \in [1 , \infty)$, such that
	\beq\label{non-purity-conditions}
	\Tr\{ \hat{\rho}^2 \} = \frac{1}{r} 
	\quad
	\text{and}
	\quad
	\sigma_{q}^2 \, \sigma_{p}^2  -  \sigma^2_{qp} = \frac{\hbar^2 \, r^2}{4} \, .
	\eeq
Consequently, for $r=1$ we have pure states and only in this case, the density operator can be factorized as $\hat{\rho} = | \alpha \rangle \langle \alpha |$, where $| \alpha \rangle$ is the so-called generalized coherent state.
In this form, the GKLS dynamics acting on the Gaussian kernel can modify the parameter $r$ introducing a change of purity. An expression for the Gaussian kernel that explicitly depends on the parameter $r$ is obtained by its Wigner representation given by the quasi-distribution function
	\beq 
	W_r( q , p ) = \frac{1}{\pi \, \hbar \, r} \exp \left\{
	- \frac{2}{\hbar^2 \, r^2} \left[
	\sigma_{p}^2 q^2 
	- 2 \sigma_{qp} q p
	+ \sigma_{q}^2 p^2
	\right]
	\right\} \, .
	\eeq
From this expression, we observe that the change in the degree of purity of the state is reflected in the form of the Wigner function, having an explicit dependence on the $r$-parameter.
To include the non-pure states in the parameter space, we consider the change of variables
	\beq 
	y^1 =  \frac{1}{\hbar} (\sigma^2_{p} + \sigma^2_{q}) \, ,
	\qquad
	y^2 =  \frac{2}{\hbar} \, \sigma_{qp} \, , 
	\qquad
	y^3 =  \frac{1}{\hbar} (\sigma^2_{p} - \sigma^2_{q}) \, ,
	\eeq	 
such that, the constrain for the second moments in \eqref{non-purity-conditions} for non-pure states in this new variables defines the space of parameters as the solid hyperboloid 
	\beq 
	\mathbf{H} = \{ (y^1, y^2, y^3) \in \R^3 \, | \, (y^1)^2 - (y^2)^2 - (y^3)^2 \geq 1 \} \, ,
	\eeq  
where the minimal surface with $r = 1$ corresponds to the hyperboloid 
	\beq 
	H^2 = \{ (y^1, y^2, y^3) \in \R^3 \, | \, (y^1)^2 - (y^2)^2 - (y^3)^2 = 1 \} \, ,
	\eeq
for pure states. 	
Therefore, there is a clear analogy between the ball $\mathbf{B}$ for the q-bit states, and the solid hyperboloid $\mathbf{H}$ for Gaussian states, in the sense that quantum states are represented as a subset of points in these finite-dimensional manifolds.
Moreover, because a Gaussian initial state remains Gaussian under evolution obtained by operators that are at most quadratic in position and momentum~\cite{Bonet-Luz-2016}, we restrict our study to observables of the form $\hat{a} = a_\mu \hat{L}^\mu$ with 
	\beq 
	\hat{L}^0 = \frac{\hbar}{2} \hat{I} \, ,
	\quad
	\hat{L}^1 = \frac{1}{4} \left( \hat{p}^2 + \hat{q}^2 \right) \, ,
	\quad
	\hat{L}^2 = \frac{1}{4} \left( \hat{q}\,\hat{p} + \hat{p}\,\hat{q} \right) \, ,
	\quad
	\text{and}
	\quad
	\hat{L}^3 = \frac{1}{4} \left( \hat{p}^2 - \hat{q}^2 \right) \, ,
	\eeq
then, the expectation value of this operator $f_{ \hat a} = \Tr\{ \hat{\rho} \, \hat{a} \} =\frac{\hbar}{4}  a_\mu y^\mu$, is a linear function on the space of parameters $\mathbf{H}$.	

The major contribution of this work is to find the GKLS vector field for Gaussian kernel, which allows a change in the degree of purity through the variation of the $r$-parameter.
Nevertheless, trying to follow the same procedure presented in \cite{Ciaglia-2017}, it is immediate to notice that the infinite representation in \eqref{Gaus-Den-Mat} for Gaussian kernel results inappropriate for such a task.
To obtain the GKLS vector field, as in the q-bit case, we consider the so-called \emph{observable point of view}.
To do this, let us note from the q-bit case that the observables are elements of a finite $C^\ast$-algebra represented in terms of $(2 \times 2 )$ matrices, which are also elements of a Lie--Jordan algebra.
Moreover, there is a notion of dual algebra, i.e., a dual vector space of the vector space with a Lie algebra structure, on which we define a Poisson tensor and a Riemann--Jordan tensor.
Then, to follow the same procedure we restrict our study to the set of observables $\hat{a} = a_\mu \hat{L}^\mu$, where the space of such observables is denoted by $\D$. 
The advantage of considering $\D$ resides in the fact that the basis $\{ \hat{L}_{\mu} \}_{\mu = 0,1,2,3}$ accepts a finite non-unitary matrix representation in terms of $2 \times 2$ matrices that close under the Lie--Jordan algebra.
In addition, if $\D^\ast$ is the dual Lie algebra, then for every $\xi \in \D^\ast$, there is a unique $\hat{\xi} \in \D$ such that
	\beq 
	\langle \xi,\hat{a} \rangle := \, \Tr \{ \hat{a} \, \hat{\xi} \} \, ,
	\eeq 
for an arbitrary observable $\hat{a} \in \D$.
In coordinates, we can define the matrix $\xi$ as
	\beq
	\xi =  \frac{1}{2 \, \hbar} \, y^\mu \, L_\mu \, ,
	\eeq
where $\mu =0,1,2,3$, $L_\mu = g_{\mu\nu} \hat{L}^\nu$ and $y^\mu \equiv g^{\mu\nu}y_\nu$, with $g^{\mu\nu}$ the entries of the matrix $\rm{diag}(1,1,-1,-1)$.
Thus, the normalization condition $\Tr\{ \hat{\xi} \} = 1$ fixes $y^0 = 2$ and the components $(y^1, y^2, y^3)$ can be obtained as
	\beq
	 \Tr\{ \hat{L}^\mu \hat{\xi} \}
	= \frac{\hbar}{4} y^\mu \, .
	\eeq
Consequently, we have that the linear function $f_{\hat{a}}$, associated with the observable $\hat{a} = a_\mu \, \hat{L}^\mu \in \D$ corresponds to 
	\beq
	f_{\hat{a}} = \langle \xi , \hat{a} \rangle = \frac{\hbar}{4} a_\mu y^\mu \, ,
	\eeq
which coincides with the expectation value of the operator.
Therefore, the matrix representation of the operator allows one to follow the same procedure proposed for the $n$-level systems presented in \cite{Ciaglia-2017}.
It is because we claim that the GKLS evolution for $\hat{\xi}$ can be determined through the master equation
	\beq
	L(\hat{\xi}) =  - \frac{\i}{\hbar} [\hat{H}, \hat{\xi}]_{-} 
	- \frac{1}{2} \sum^3_j [ \hat{v}^\dagger_j \, \hat{v}_j , \hat{\xi} ]_{+}
	+ \sum^3_j \hat{v}^\dagger_j \, \hat{\xi} \, \hat{v}_j \, ,
	\eeq
where $\hat{H}$ and $\hat{v}$ are elements of $\D$ and, consequently, they can be represented by $(2 \times 2)$-matrices.
Therefore, we can follow the same procedure for the q-bit system considered in Ref.~\cite{Ciaglia-2017}, up to the appropriate modifications, to obtain the GKLS vector field for the Gaussian states.

The paper is organized as follows. In Section \ref{sec.2} we review the case of one q-bit systems, starting with the description of the space of quantum states as a manifold with boundary and establishing its kinematic properties, in particular, we describe its foliation in terms of two-spheres. In Subsection \ref{subsec.2.1}, we describe the Hamiltonian dynamics on each isospectral submanifold through its K\"ahler structure. 
In Subsection \ref{subsec.2.2}, we analyze the quantum systems from the \emph{point of view of observables} and use the Lie-Jordan algebra to define the two relevant geometric structures, a skew-symmetric bivector field which defines a Poisson structure for the space of functions on the dual algebra, that is associated to the Lie product and a symmetric bivector field which defines a symmetric product, both are realizations of the algebra on the space of linear functions. 
In Subsection \ref{subsection.2.3}, we construct the GKLS vector field and find the decomposition  \eqref{Gamma-n-level}, to do so, we perform a reduction procedure and find the Choi-Kraus vector field  associated to the completely positive map $\mathcal{K}$ in \eqref{GKLS-generator}. 
In the last subsection, \ref{subsec.2.4}, we present two examples for a damping process of a two-level system. 
In Section \ref{sec.3}, and its subsections, our main results are presented, where the procedure reviewed in Section \ref{sec.2}, with the appropriate tuning, is used to determine the GKLS vector field for the Gaussian states. 
It is worth to mention that, adapting the procedure to an infinite-dimensional state space is not straightforward, and some modifications had to be considered.
Finally, in Section \ref{sec.4}, we present our conclusions and some of the lines of research that will be presented in a set of future works.


\section{ On the kinematics and dynamics of Dissipative one $q$-bit systems} \label{sec.2}

In this section, the GKLS dynamical vector field for one $q$-bit systems is constructed in detail to exemplify the ideas that we will apply in section \ref{sec.3} for the Gaussian density matrices.
It is well known that in general, the quantum space for the $q$-bit can be immersed in the space of $2 \times 2$ Hermitian matrices, where the basis may be provided by the Pauli matrices
	\beq \label{algebra-basis}
	\sigma_1 = \left(
	\begin{array}{c c}
	  0 &  1\\
	  1 & 0
	\end{array}
	\right) \, , 
	\quad
	\sigma_2 = \left(
	\begin{array}{c c}
	  0 & -\i \\
	  \i  & 0
	\end{array}
	\right) \, ,
	\quad
	\sigma_3 = \left(
	\begin{array}{c c}
	  1 &  0\\
	  0 & -1
	\end{array}
	\right) \, ,
	\eeq
and the identity matrix
	\beq \label{Identity}
	\sigma_0  = \left(
	\begin{array}{c c}
	  1 &  0\\
	  0 & 1
	\end{array}
	\right) \, .
	\eeq
Thus, an arbitrary density matrix may be expressed as \footnote{ It is important mention that here, and in the following, $\{ \sigma_{\mu} \}_{\mu=0,1,2,3}$ denotes the basis for the space $\D_2^\ast$, which is dual to the basis $\{ \hat{\sigma}^{\mu} \}_{\mu=0,1,2,3}$ for $\D_2$, defined in the introduction as the set of self-adjoint operators. 
This distinction between elements of the algebra and its dual will be highlighted by using a hat to denote operators.} 
	\beq \label{q-rho}
	\rho = \frac{1}{2} (\sigma_0 + x^k \sigma_k) = \frac{1}{2} \, x^\mu \, \sigma_\mu \, ,
	\eeq
where $k = 1,2,3$, $\mu =0,1,2,3$ and from the normalization condition $\Tr\{ \hat{\rho} \, \hat{\sigma}^0 \} = 1$ it follows that $x^0 = 1$. 
In this expression, and from here on, Einstein's summation convention over repeated indices is assumed\footnote{Throughout the paper greek indices will run from 0 to 3, meanwhile latin from 1 to 3.}.
Thus, for example, every pure quantum state is represented by a point $(x^1, x^2, x^3)$ in the unit sphere, such that $x^k~=~\Tr\{ \hat{\sigma}^k \hat{\rho} \}$. 
Here the purity condition $\rho^2 = \rho$ defines the unit sphere
	\beq
	S^2 = \{ (x^1, x^2, x^3) \in \R^3 \, | \, (x^1)^2 + (x^2)^2 + (x^3)^2 = 1 \} \, ,
	\eeq
which is known as the Bloch sphere.
On the other hand, for mixed states one has that the mixture condition $\Tr\{ \hat{\rho}^2 \} < 1$ defines the constraint
	\beq
	(x^1)^2 + (x^2)^2 + (x^3)^2 < 1 \, ,
	\eeq
where the maximal mixed state correspond to $x^1 = x^2 = x^3 = 0$.
Therefore, a $q$-bit state may be always represented by a point $\mathbf{x} = (x^1, x^2 , x^3)$ on the solid ball
	\beq \label{Ball}
	\mathbf{B} = \{ (x^1, x^2, x^3) \in \R^3 \, | \, (x^1)^2 + (x^2)^2 + (x^3)^2 \leq 1 \} \, .
	\eeq
In the literature, the points $\mathbf{x} \in \mathbf{B}$ are called \emph{Bloch vectors} or \emph{polarization vectors}~\cite{Scully-1999}.
The space of 2-level quantum system is made up of two strata: $\mathcal{S}_1$ by unit sphere $S^2$, the space of pure states and $\mathcal{S}_2$ the open interior of the ball, space of mixed states.
Therefore, the quantum space of q-bit systems is a manifold with boundary.

\begin{figure}[! t]
	\centering
	\includegraphics[width = 8 cm]{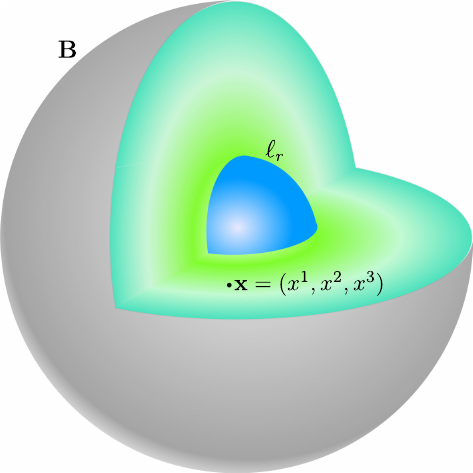}
	\caption{Pictorial representation of the space of quantum states for the $q$-bit systems. 
	A quantum state is represented by a point $\mathbf{x} = (x^1, x^2, x^3)$ in the ball $\mathbf{B}$.
	In addition, one may see that the ball is foliated by a disjoint family of spheres $\ell_r$. }
	\label{Fig-1}
	\end{figure}

As a final remark, let us notice that the manifold $\mathbf{B}$ is a foliated space, i.e.,
	\beq
	\mathbf{B} = \bigcup_{r \in [0 , 1]} \ell_r
	\eeq
where the leaves of the foliation correspond to 
	\beq \label{Spheres}
	\ell_r = \{ (x^1, x^2, x^3) \in \R^3 \, | \, (x^1)^2 + (x^2)^2 + (x^3)^2 = r^2, r \leq 1 \} \, .
	\eeq
A schematic picture of this foliation is displayed in Fig. \ref{Fig-1}.
It is important to note that we have a \emph{singular} foliation, i.e., the leaves are not all of the same dimension, having a singular point at the origin. 
Nevertheless, removing the origin one has a regular foliation given by the family $\{ \ell_r \}$ of disjoint subsets, with $r \in (0 , 1]$ and where $\ell_r$ are the leaves of the foliation, on which a differential structure can be given. 
In general, this foliation is a consequence of the smooth action of $SU(\a_2)$.
In the literature, the leaves of this foliation are the so-called \emph{manifolds of isospectral states}~\cite{Ciaglia-2017, Chruscinski-2019}.
	
\subsection{Dynamical study of one q-bit systems from a state point of view}
 \label{subsec.2.1}

Once we have introduced the manifolds of isospectal states, in this section we proceed to define the symplectic dynamics and the gradient vector field on these manifolds.
To do that, let us note that each manifold of isospectral states is endowed with a \emph{K\"ahler structure}, i.e., there is a symplectic form $\omega_{\ell_r}$, a Riemannian metric $g_{\ell_r}$, and a complex structure $J_{\ell_r}$ all of them defined globally on each isospectral manifold~\cite{Grabowski-2005, Ercolessi-2010, Cruz-2020}. 
For $r = 1$ the Riemannian metric is known as \emph{the Fubini--Study metric}. 
Moreover, the symplectic form and the Riemannian metric define a Hamiltonian vector field $\mathbb{X}_H$ and a gradient vector field $\mathbb{Y}_H$. 
Given a real function $f_{\hat H} \in \D^\ast_2$ defined as the expectation value of the observable $\hat{H} \in \D_2$, that is,  
	\beq
	f_{\hat H} := \Tr\{ \hat{\rho} \, \hat{H} \} \, ,
	\eeq
then the Hamiltonian vector field and the gradient vector field are defined intrinsically by
	\beq \label{Isospectal-fields}
	\omega_{\ell_r}( \mathbb{X}_{H} , \, \cdot \, ) = \d f_{\hat H}
	\quad
	\text{and}
	\quad
	g_{\ell_r} ( \mathbb{Y}_{H} , \, \cdot \, ) = \d f_{\hat H} \, ,
	\eeq
respectively, along with the property
	\beq \label{Comp-Struc-Qbit}
	J_{\ell_r}( \mathbb{X}_H ) = \mathbb{Y}_{H} \, ,
	\eeq
which provides an intrinsic definition of the complex structure tensor $J_{\ell_r}$.

To give the coordinate expressions for all these definitions, let us consider the coordinate charts $(U_1, \phi_1)$, $(U_2, \phi_2)$ for the foliation $\ell_r$, with $U_1, U_2 \subset \ell_r$ and
	\beq
	\phi_1 : U_1 \to \C : (x^1, x^2, x^3) \mapsto z
	\quad
	,
	\quad
	\phi_2 : U_2 \to \C : (x^1, x^2, x^3) \mapsto \zeta \, ,
	\eeq
where the complex parameters $z$ and $\zeta$ are given by
	\beq \label{Ric-Rect}
	z = \frac{x^1 - \i x^2}{r - x^3} \, ,
	\qquad
	\zeta = \frac{x^1 + \i x^2}{r + x^3} \, .
	\eeq
In this manner, the set $\{ (U_1, \phi_1),  (U_2 , \phi_2) \}$ constitutes an atlas for each foliation $\ell_r$.
These coordinates in quantum mechanics are employed to describe \emph{atomic coherent states}~\cite{Arecchi-1972} or \emph{spin coherent states} \cite{Radcliffe-1971}, see also \cite{Perelomov-1977, Zhang-1990}.

Geometrically, the atlas $\{ (U_1, \phi_1), (U_2, \phi_2) \}$ corresponds to the stereographic projection from the ``north pole'' and the ``south pole'' of the sphere onto the equatorial plane, respectively. 
Notice that, using the Cartesian coordinates $(x^1, x^2, x^3)$ we are giving an extrinsic geometric description of the system, in which case one obtains a set of linear equations of motion. 
On the other hand, the stereographic projection atlas constitutes an intrinsic geometric description and, as we will see, the equations of motions are non-linear.

Now, in the coordinate chart $(U_1, \phi_1)$ the symplectic form and the Riemannian metric are given by
	\beq \label{forms}
	\omega_{\ell_r} = \frac{ - \i \, \hbar \, r^2}{(1 + \bar{z} z )^2} \d \bar{z} \, \wedge \, \d z \, ,
	\quad
	\text{and}
	\quad
	g_{\ell_r} =  \frac{ - \hbar \, r^2}{(1 + \bar{z} z )^2} \d \bar{z} \, \otimes_{\tiny \mbox{s}} \, \d z \, ,
	\eeq
respectively. 
In these definitions one considers the wedge product $\d \bar{z} \, \wedge \, \d z = \d \bar{z} \otimes \d z - \d z \otimes \d \bar{z}$ together with the symmetrical product $\d \bar{z} \, \otimes_{\tiny \mbox{s}} \, \d z = \d \bar{z} \otimes \d z + \d z \otimes \d \bar{z}$, while the complex structure has the form
	\beq
	J_{\ell_r} = \frac{1}{\i}
	\left(
	\d z \otimes \frac{\partial}{\partial z } - \d  \bar{z} \otimes \frac{\partial}{\partial  \bar{z} }
	\right) \, .
	\eeq
To express the density matrix \eqref{q-rho}  in this coordinate system, one has to take into account the inverse of the stereographic projection to obtain the relations
	\beq \label{relation-x-z}
	x^1 = r \frac{z + \bar{z}}{1 + z \bar{z}} \, ,
	\quad
	x^2 = - \frac{r}{\i}\frac{z - \bar{z}}{1 + z \bar{z}}
	\quad
	\text{and}
	\quad
	x^3 = r \frac{z\bar{z} - 1}{1 + z \bar{z}} \, ,
	\eeq
and then, by direct substitution, the density matrix has the form
	\beq \label{rho-z}
	\rho = \frac{1}{1 + \bar{z} z } \left(
	\begin{array}{ccc}
	\frac{1}{2}(1 - r) + \frac{1}{2}( 1 + r )\bar{z} z & r \, z \\
	& \\
	r \, \bar{z} & \frac{1}{2}(1 + r) + \frac{1}{2}( 1 - r )\bar{z} z \\   
	\end{array}
	\right) \, ,
	\eeq
where the dependence on the $r$-parameter is explicit.
From the density matrix in Eq. \eqref{rho-z}, the expectation value of an arbitrary observable operator can be obtained.
In general, a self-adjoint operator can be written as $\hat{H} = H_\mu \, \hat{\sigma}^\mu$; thus, its expectation value corresponds to 
	\beq
	f_{\hat H}  := \Tr\{ \hat{\rho} \, \hat{H} \} = H_0 + H_k x^k \, , 
	\eeq 
and taking into account the transformations in \eqref{relation-x-z}, it can be expressed as 
	\beq
	f_{\hat H} = \frac{1}{1 + \bar{z}z}
	\left[ ( H_0 + r \, H_3 ) z \bar{z} + r (H_1 + \i \, H_2) z + r (H_1 - \i \, H_2) \bar{z}
	+ ( H_0 - r \, H_3) \right] \, ,
	\eeq
where $f_{\hat H}$ shows the explicit dependence on the parameter $r$, while the Hamiltonian and gradient vector fields take the form
	\beq \label{Vector-Fields}
	\mathbb{X}_{H} = \mathbb{X}_z \frac{\partial}{\partial z} 
	+ \mathbb{X}_{\bar{z}} \frac{\partial}{\partial \bar{z}}
	\quad
	\text{and}
	\quad
	\mathbb{Y}_{H} = \mathbb{Y}_z \frac{\partial}{\partial z} 
	+ \mathbb{Y}_{\bar{z}} \frac{\partial}{\partial \bar{z}} \, ,
	\eeq
where the components $\mathbb{X}_{\bar{z}}$ and $\mathbb{Y}_{\bar{z}}$ are the complex conjugated of $\mathbb{X}_z$ and $\mathbb{Y}_z$, respectively.
These components may be directly computed by means of the definitions in Eq.~\eqref{Isospectal-fields} to obtain that
	\beq
	\mathbb{X}_z = \frac{- \i}{\hbar \, r^2} (1 + \bar{z} z)^2 \frac{\partial f_{\hat H}}{\partial \bar{z}}
	= \frac{\i}{\hbar \, r} \left[ (H_1 + \i H_2) z^2 - 2 H_3 \, z - (H_1 - \i H_2) \right] \, ,
	\eeq
and
	\beq \label{Gradient-component}
	\mathbb{Y}_z =  \frac{- 1}{\hbar \, r^2} (1 + \bar{z} z)^2 \frac{\partial f_{\hat H}}{\partial \bar{z}}
	= \frac{1}{\hbar \, r} \left[ (H_1 + \i H_2) z^2 - 2 H_3 \, z - (H_1 - \i H_2) \right] \, .
	\eeq
Then, an important consequence of giving an intrinsic description of the manifolds of isospectral states is having a non-linear evolution equation, in this case we have obtain the \emph{non-linear Riccati equation}
	\beq \label{Qbit-Ric-Eq}
	\dot{z} = \frac{\i}{r \, \hbar}  \left[ (H_1 + \i H_2) z^2 - 2 H_3 \, z - (H_1 - \i H_2) \right] \, ,
	\eeq
as a Hamiltonian evolution for the quantum states; besides, note that this evolution is independent on $H_0$.
On the other hand, the equations of motion in the extrinsic geometric description with coordinates $(x^1,x^2, x^3)$ are given by the following system of differential equations
	\begin{align}
	\dot{x}^1 & = - \frac{2}{\hbar \, r} (x^2 H_3 - x^3 H_2) \, , \nonumber \\
	\dot{x}^2 & = - \frac{2}{\hbar \, r} (x^3 H_1 - x^1 H_3) \, ,  \nonumber \\
	\dot{x}^3 & = -  \frac{2}{\hbar \, r} (x^1 H_2 - x^2 H_1) \, ,
	\end{align}
and whose solutions correspond to the integral curves of the Hamiltonian vector field
	\beq \label{XH-Cartesian-Coordinates}
	\mathbb{X}_{H} = - \frac{2}{\hbar \, r} \epsilon^{k j}_l x^l \, H_k \frac{\partial}{\partial x^j} \, ,
	\eeq	
where $ \epsilon^{kj}_l$ is the Levi-Civita symbol\footnote{The convention for the Levi-Civita symbol is the following
	$$
	\epsilon^{kj}_l 
	= \begin{cases}
	1  & \text{if} \quad (k,j,l) \quad \text{is} \quad (1,2,3), (2,3,1), (3,1,2) \, , \\
	-1 & \text{if} \quad (k,j,l)\quad \text{is} \quad (2,1,3), (3,2,1), (1,3,2) \, , \\
	0  & \text{if} \quad k=j, \quad j=l, \quad l=k
	\end{cases}
	$$}.
Let us note that in the above set of equations, the parameter $r$ is a constant of the motion; therefore, its value becomes fixed once the initial conditions are given, and it is only then that the equations are linear. Hence, the coefficients of the matrix associated to the linear system of equations change as we choose different initial conditions.

Alternatively, the evolution of the quantum systems can be described directly by the so-called Poisson brackets and Jordan brackets on the manifolds of isospectral states~\cite{Grabowski-2005, Ercolessi-2010, Cruz-2020}.
Given the expectation values $f_{\hat a}$ and $f_{\hat b}$ associated to the quantum observables $\hat{a}$ and $\hat{b}$, one can define the Poisson brackets and Jordan brackets through the relations
	\beq \label{Braquets}
	\{ f_{\hat a} , f_{\hat b} \}_{\omega} 
	= \omega_{\tiny\mbox{FS}}(X_b , X_a ) \, , 
	\qquad
	\langle \langle f_{\hat a} , f_{\hat b} \rangle\rangle_{g} 
	= g_{\tiny\mbox{FS}}(Y_a, Y_b ) \, ,
	\eeq
respectively, and where these brackets satisfy the relations\footnote{Notice that in the definition of Hamiltonian vector field we are using the convention
	$$
	\omega(X_H, \cdot) = \d H\, ,
	$$ 
which has a minus sign with respect to the more common choice in Classical Mechanics, i.e.,
	$$
	\omega(X_H, \cdot) = - \d H\, ,
	$$ 
as is taken, for instance, in \cite{Arnold-2013}, but following the same definition for the Poisson bracket from the symplectic form $$\{f, g\} = \omega(g, f) = -\omega(f,g)\, .$$ }
	\beq \label{Brakets-commutators}
	\{ f_{\hat a}, f_{\hat b} \}_{\omega} 
	= - \frac{1}{r} f_{[[ \hat{a}, \hat{b} ]]}
	\quad
	\mbox{and}
	\quad
	\langle \langle f_{\hat a} , f_{\hat b} \rangle\rangle_{g} 
	= - \, f_{\hat{a} \odot \hat{b}} + \frac{2}{\hbar \, r^2} \, f_{\hat{a}} \, f_{\hat{b}} \, ,
	\eeq
with
	\beq \label{Lie-Jordan-algebra}
	[[ \hat{a}, \hat{b} ]] = \frac{\i}{\hbar}( \hat{a} \, \hat{b} - \hat{b} \, \hat{a} )
	\quad
	\text{and}
	\quad	
	\hat{a} \odot \hat{b} = \frac{1}{\hbar}( \hat{a} \, \hat{b} + \hat{b} \, \hat{a} ) \, ,
	\eeq	
defining the Lie and Jordan products, respectively. 
Using the Poisson and the Jordan brackets means that we are describing the quantum systems from \emph{the point of view of observables $f_{\hat a}$ as the primary objects, hence states and dynamics are derived from it}. 
The observables in quantum mechanics constitute a Lie--Jordan algebra $(\D_n, \odot, [[\cdot,\cdot]])$ where the products satisfy the following compatibility conditions
	\beq
	[[ \hat{a}, \hat{b} \odot \hat{c} ]] = [[ \hat{a}, \hat{b} ]] \odot \hat{c} + \hat{b} \odot [[ \hat{a},\hat{c} ]] \, ,
	\eeq
and
	\beq
	(\hat{a} \odot \hat{b}) \odot \hat{c} - \hat{a} \odot (\hat{b} \odot \hat{c})
	= [[ \hat{b}, [[ \hat{c} , \hat{a}]] ]]  \, .
	\eeq	
Thus, for the Lie algebra structure, observables appear as infinitesimal generators of one-parameter groups of transformations and with respect to the Jordan structure observables appear as measurable quantities with outcomes given by real numbers, more specifically, probability measures on the real line. 

Including $r$ as a dynamic variable means that the differential manifold of quantum states is odd-dimensional, therefore 
a symplectic form cannot defined.
Nevertheless, this alternative description in terms of a Poisson structure allows to extend our definition of the Hamiltonian vector field to all of $\mathbf{B}$. 
Furthermore, as we will see in the following section, from this perspective we have a direct route to the GKLS dynamics.
	
To conclude this subsection, let us introduce two important geometrical objects: the skew-symmetrical bivector $\Lambda_{\ell_r}$ and the symmetrical bivector $G_{\ell_r}$, which in terms of the stereographic coordinates take the form
	\beq \label{Conservative-Bivector}
	\Lambda_{\ell_r} = - \frac{\i}{r^2 \, \hbar}(1 + \bar{z} z)^2 
	\frac{\partial}{\partial \bar{z}} \wedge \frac{\partial}{\partial z}
	\quad
	\text{and}
	\quad
	G_{\ell_r} = - \frac{1}{r^2 \, \hbar}(1 + \bar{z} z)^2 
	\frac{\partial}{\partial \bar{z}} \otimes_{\tiny \mbox{s}} \frac{\partial}{\partial z} \, ,
	\eeq
such that the Poisson and the Jordan brackets may be defined as
	\beq
	\{ f_{\hat a}, f_{\hat b} \}_{\omega} 
	= \Lambda_{\ell_r} ( \d f_{\hat a} , \d f_{\hat b})
	\quad
	\mbox{and}
	\quad
	\langle \langle f_{\hat a}, f_{\hat b} \rangle \rangle_{g} 
	= G_{\ell_r} ( \d f_{\hat a} , \d f_{\hat b} ) \, ,
	\eeq
respectively. 
These geometrical objects will be relevant in the GKLS evolution.


\subsection{Dynamical study of one q-bit systems from an observable point of view} \label{subsec.2.2}

The vector fields in \eqref{Vector-Fields}, for a given $r$ determined by the initial conditions, are tangent to the manifolds of isospectral states, i.e., the quantum states resulting from the evolution of states with initial conditions on $\ell_r$, remain on such manifolds; however, the dynamical evolution given by the GKLS master equation changes the rank and the spectrum of quantum states, i.e., the description of dissipative phenomena leads to consider an evolution that is transversal to the leaves $\{ \ell_r \}$.
It is important to mention that the dynamical evolution for any set of initial conditions must be constrained to the space of quantum states.
The evolution of a quantum state determined by the GKLS master equation is non-unitary, completely positive and trace preserving.

As it was mentioned in the introduction, the geometrical formulation of the dynamics of open quantum systems generated by the GKLS equation is given by an affine vector field $\Gamma$, which may be decomposed as
	\beq \label{Affine-vector-field}
	\Gamma = X_H + Y_b + Z_{\mathcal K},
	\eeq  
where $X_H$ is a Hamiltonian vector field on $\mathbf{B}$ which describe a conservative dynamics and the term $Y_b + Z_K$ is a perturbation term, that determines the dissipative part of the dynamics.
The two latter vector fields produce different effects in the dynamical evolution of the quantum states. 
On the one hand, $Y_b$ is a gradient-like vector field whose flow changes the spectrum but preserves the rank of the density matrix $\rho$; on the other hand, $Z_{\mathcal K}$ is responsable for the change of rank, that is, only through $Z_{\mathcal K}$ the statistical mixture of the initial state can change.

To determine the vector field $\Gamma$ we need to extend the Hamiltonian vector field to the space $\mathbf{B}$ and introduce the concept of gradient-like vector field. 
We will adopt the \emph{point of view of observables} as the primary objects from which the dynamics is obtained. 
As a starting point, we establish the Lie--Jordan algebra structure $(\D_2 , \odot, [[\cdot,\cdot]])$ of the space of observables from the Lie and Jordan products in $\D_2$. 

To construct the vector field $\Gamma$, let us start extending the definition of the Hamiltonian vector field to all $\mathbf{B}$ and introduce the concept of gradient-like vector field.
From the observable point of view, one must start employing the Lie--Jordan algebra structure $(\D_2 , \odot, [[\cdot,\cdot]])$ of the space of observables. 
To every element $\hat{a} \in \D_2$ correspond a linear function $\tilde{f}_{\hat a}$ in $\D_2^\ast$ by 
	\beq  \label{map}
	\tilde{f}_{\hat{a}}(\rho) := \Tr \{\hat{a} \, \hat{\rho} \},
	\eeq
where $\hat{\rho}$ is the density matrix.
Conversely, any linear function $\tilde{f}_{\hat a} \in \D^\ast_2$ maps to an element $\hat{a} \in \D_2$. 
Therefore, the space of linear functions on $\D^\ast_2$ together with the products defined as
	\beq \label{Lie-Jordan-realizations}
	\{ \tilde{f}_{\hat a}, \tilde{f}_{\hat b} \} = \tilde{f}_{[[ \hat{a}, \hat{b} ]]}
	\qquad
	\mbox{and}
	\qquad
	\langle \langle \tilde{f}_{\hat a}, \tilde{f}_{\hat b} \rangle \rangle= \tilde{f}_{\hat{a} \odot \hat{b}} \, .
	\eeq	
constitutes a realization of the Lie-Jordan algebra $(\D_2,\odot, [[\cdot, \cdot]])$.
	
From this algebraic structure, it is possible to define symmetric and skew-symmetric covariant tensor fields (bivectors) on $\D^\ast_2$. 
The skew-symmetric bivector $\tilde{\Lambda}$ and the symmetric bivector $\tilde{G}$ are uniquely determined by their action on the one-forms $\d \tilde{f}_{\hat a}$, which at each point in $\D^\ast_2$ are elements of the cotangent space, i.e., 
	\beq \label{Pois-Bivector}
	\tilde{\Lambda}(\d \tilde{f}_{\hat a}, \d \tilde{f}_{\hat b}) := \tilde{f}_{ [[\hat{a} , \hat{b}]] }
	\qquad
	\text{and} 
	\qquad
	\tilde{G}(\d \tilde{f}_{\hat a}, \d \tilde{f}_{\hat b}) := \tilde{f}_{\hat{a} \odot \hat{b}} \, .
	\eeq
Hence, Hamiltonian and gradient-like vector fields can be defined as
	\beq \label{GL-relization}
	\tilde{X}_H = \tilde{\Lambda}( \d \tilde{f}_{\hat H}, \, \cdot \, )
	\qquad
	\text{and}
	\qquad
	\tilde{Y}_b = \tilde{G}( \d \tilde{f}_{\hat b}, \, \cdot \,) \, ,
	\eeq
respectively. 
Notice that the bivectors $\tilde{\Lambda}$ and $\tilde{G}$ in \eqref{Pois-Bivector} are different to the bivectors \textcolor{magenta}{$\Lambda_{\ell_r}$} and \textcolor{magenta}{$G_{\ell_r}$} defined in \eqref{Conservative-Bivector}, the latter are defined on a single isospectral manifold with fixed $r$, while $\tilde{\Lambda}$ and $\tilde{G}$ are defined for any linear function in the dual space $\D^\ast_2$ with $r$ variable.
This procedure allows to define a Hamiltonian vector field $\tilde{X}_H$ by means of the Poisson bivector  $\tilde{\Lambda}$ deduced from the Lie algebra structure of $\D_2$, which is less restrictive and more general than the definition of Hamiltonian vector field in terms of the symplectic form. 
Moreover, the Jordan product allows to introduce the gradient-like vector field $\tilde{Y}_b$. 
We can introduce Cartesian coordinates $\{x^\mu\}$ associated to the basis $\{\hat{\sigma}^\mu \}$ of $\D_2$ by the mapping \eqref{map}, then the coordinate functions on $\D^\ast_2$ are
	\beq \label{cartesian-coordinates}
	x^\mu := f_{\hat{\sigma}^\mu} = \Tr\{ \hat{\sigma}^\mu \hat{\rho} \} \, . 
	\eeq
The tensor fields \eqref{Pois-Bivector} in this coordinate basis are
	\beq
	\tilde{\Lambda} = c_\eta^{\mu\nu} x^\eta \frac{\partial}{\partial x^\mu} \wedge \frac{\partial}{\partial x^\nu}
	\qquad
	\text{and} 
	\qquad
	\tilde{G} = d_\eta^{\mu\nu} x^\eta \frac{\partial}{\partial x^\mu} \otimes \frac{\partial}{\partial x^\nu} \, ,
	\eeq
where the structure constants $c_\eta^{\mu\nu}$ and $d_\eta^{\mu\nu}$ are defined uniquely by the Lie and the Jordan products, i.e., for the Lie product we have 
	\beq
	[[\hat{\sigma}^\mu, \hat{\sigma}^\nu ]] = c_\eta^{\mu\nu} \hat{\sigma}^\eta \quad \text{where} \quad
	c_\eta^{\mu\nu} = \begin{cases}
	0  & \text{for} \quad c_\eta^{0 \nu}, \, \, c_\eta^{\mu 0}, \, \, c_0^{\mu\nu} \, , \\
	\\
	- \frac{2}{\hbar} \epsilon^{kj}_l & \text{for} \quad k, j, l = 1,2,3 \, ,
	\end{cases}
	\eeq
where $ \epsilon^{kj}_l$ is the Levi-Civita symbol. 
For the Jordan product $\hat{\sigma}^\mu \odot \hat{\sigma}^\nu = d_\eta^{\mu\nu} \hat{\sigma}^\eta$ where 
	\beq
	\hat{\sigma}^\mu \odot \hat{\sigma}^\nu = d_\eta^{\mu\nu} \hat{\sigma}^\eta 
	\quad 
	\text{with} 
	\quad
	d_\eta^{\mu\nu} = \begin{cases}
	0  & \text{for} \quad d_l^{0 0} \quad \text{and} \quad d_l^{k j} \, , \\
	\\
	\frac{2}{\hbar} \delta^{\mu \nu} & \text{for} \quad d_0^{\mu\nu} \, , \\
	\\
	\frac{2}{\hbar} \delta^{\mu}_\eta & \text{for} \quad d_\eta^{\mu 0} = d_\eta^{0 \mu} \, ,
	\end{cases}
	\eeq 
where $\delta^{\mu \nu}$ and $\delta^{\mu}_\eta$ are Kronecker delta functions. Let us now compute the coordinate expressions for the Hamiltonian and the gradient-like vector fields using the definitions in \eqref{GL-relization}, that is, considering the expectation values $\tilde{f}_{\hat{H}} = H_\mu \, x^\mu$ and $\tilde{f}_{\hat{b}} = b_\mu \, x^\mu$ we obtain	
	\beq \label{Vector-fields}
	 \tilde{X}_{H} = c_\sigma^{\mu\nu} H_\mu x^\sigma \frac{\partial}{\partial x^\nu}
	 \qquad
	\text{and} 
	\qquad
	\tilde{Y}_b = d_\sigma^{\mu\nu} b_\mu x^\sigma \frac{\partial}{\partial x^\nu} \, ,	
	\eeq
respectively.

Because the dynamical trajectories of quantum states under the action of $\Gamma$, in Eq. \eqref{Affine-vector-field}, must remain in the space of physical states, then, it is necessary to constraint further the manifold where the Hamiltonian $\tilde{\Lambda}$ and gradient-like $\tilde{G}$ vector fields act. Thus, we must consider the affine subspace
	\beq \label{Affine-space}
	\T_2^1 = \{ \xi \in \D_2^\ast \, | \, \Tr\{ \hat{\xi} \, \mathbb{I} \} = 1 \} \, ,
	\eeq 
that is, all those elements in $\D_2^\ast$ with $x^0 = 1$.
This fact allows to introduce the canonical immersion $i:= \T_2^1 \to \D_2^\ast$, such that the pullback $i^\ast \tilde{f}_{\hat{a}} = f_{\hat{a}} = a_0 + a_k x^k$ of a linear function $\tilde{f}_{\hat{a}} = a_\mu x^\mu$ associated to $\hat{a} \in \D_2$ is an affine function on $\T_2^1$.
Consequently, we can define symmetric and skew-symmetric tensor fields $\Lambda$ and $G$ on $\T_2^1$ through a reduction procedure for the bivectors $ \tilde{\Lambda}$ and $ \tilde{G}$.

Then, as has been pointed out in Ref.~\cite{Ciaglia-2017}, the algebra $\mathcal{F}(\T_2^1)$ of functions on the affine space $\T_2^1$ may be identified with the quotient space $\mathcal{F}(\D_2^\ast)/\mathcal{I}$, where $\mathcal{F}(\D_2^\ast)$ is the algebra of smooth functions in $\D_2^\ast$ and $\mathcal{I}$ is the closed linear subspace of smooth functions vanishing on the affine space.

The quotient space $\mathcal{F}(\D_2^\ast)/\mathcal{I} \cong \mathcal{F}(\T_2^1)$ inherits the vector space structure from $\mathcal{F}(\D_2^\ast)$. 
For $\mathcal{F}(\T_2^1)$ also inherit the algebra structure with respect to the relevant algebraic product, the subspace $\mathcal{I}$ must be an ideal of $\mathcal{F}(\D^\ast_2)$.
For the Poisson product the reduction is straightforward. 
Considering $\tilde{f}_{\hat{b}} = (1 - x^0) b_k x^k$ in $\mathcal{I}$, it can be shown by direct calculation that for an arbitrary $\tilde{f}_{\hat a} \in \mathcal{F}(\D^\ast_2)$, the realization of the Poisson bracket through the bivector $\tilde{\Lambda}$ gives
	\beq
	\tilde{\Lambda}(\d \tilde{f}_{\hat a}, \d \tilde{f}_{\hat b} ) = (1-x^0) \, c^{k l}_j \, x^l(a_k b_l -b_k a_l)  \ .
	\eeq 
Thus, $\tilde{\Lambda}(\d \tilde{f}_{\hat a}, \d \tilde{f}_{\hat b} )$ vanishes on $\T_2^1$, meaning that the Poisson product defined by \eqref{Lie-Jordan-realizations} and \eqref{Pois-Bivector} is such that $\{\tilde{f}_{\hat b}, \tilde{f}_{\hat a}\} \in \mathcal{I}$ for $\tilde{f}_{\hat b} \in \mathcal{I}$, therefore, $\mathcal{I}$ is an ideal of $\mathcal{F}(\D^\ast_2)$. 
Then, reducing the bivector field $\tilde{\Lambda}$ we can find a bivector field $\Lambda$ that permits to define the Poisson product on the affine space $\T_1^2$.

To perform the reduction we choose $x^k = f_{\hat{\sigma}^k}$ as a basis of $\T_2^1$, then their differentials form a basis for the cotangent space at each point of $\T_2^1$; then, through the pullback induced from the immersion $i$ we have the reduced bivector $\Lambda$ in $\T_2^1$ given by
	\beq
	\Lambda(\d f_{\hat{\sigma}^k}, \d f_{\hat{\sigma}^l}) := f_{ [[ \hat{\sigma}^k , \hat{\sigma}^l]] } 
	= c^{kl}_j f_{\hat{\sigma}^j} \, .
	\eeq 
Thus, the explicit expression of $\Lambda$ in Cartesian coordinates is 
	\beq \label{Poisson-Bivector}
	\Lambda = - \frac{2}{\hbar} \, \epsilon^{kj}_l  x^l \frac{\partial}{\partial x^k} \wedge \frac{\partial}{\partial x^j} \, .
	\eeq
Using this Poisson bivector $\Lambda$, it is posible to define the Hamiltonian vector field associated with the linear function $f_{\hat H} = H_0 + H_k x^k$ by 
	\beq
	X_H := \Lambda( \d f_{\hat H} , \, \cdot \, ) 
	= - \frac{2}{\hbar} \, \epsilon^{kj}_l  H_k x^l \frac{\partial}{\partial x^j} \, .
	\eeq 	
In addition, it is direct to note that the function 
	\beq
	f_{\hat C} = (f_{\hat{\sigma}^1})^2 + (f_{\hat{\sigma}^2})^2 + (f_{\hat{\sigma}^3})^2 = (x^1)^2 + (x^2)^2 + (x^3)^2 
	\eeq
is a constant of motion as $\Lambda( \d f_{\hat H} , \d f_{ \hat C}) = 0$.
This implies that the Hamiltonian vector field is tangent to the spheres defined by $f_{\hat C} = r^2$.
In this case, the affine space $\T_2^1$ actually corresponds to the foliated ball $\mathbf{B}$, already defined in Eq. \eqref{Ball}, whose center has been fixed at the origin of the Cartesian space $\R^3$.

The gradient-like vector field $\tilde{Y}_b$ defined on $\D^\ast_2$ from the bivector $\tilde{G}$ in \eqref{GL-relization} has integral curves that do not preserve the trace of $\xi$, therefore, starting with initial conditions on $\T_2 ^1$ the dynamics provided by $\tilde{Y}_b$ may lead to non-physical states $\mathrm{Tr}\{\hat{\xi}\, \mathbb{I}\} \neq 1$. 
In particular, starting on the boundary of $\T_2^1$ (the boundary of $\mathbf{B}$) the integral curves may end up outside of $\mathbf{B}$. 
In order to avoid this behaviour, we proceed to perform the reduction of the symmetrical bivector $\tilde{G}$.
For $\tilde{f}_{\hat a} = a_\mu x^\mu \in \D^\ast_2$ and $\tilde{f}_b =(1 - x^0) b_k x^k  \in \mathcal{I}$, it is not difficult to find that
	\beq
	\tilde{G}(\d \tilde{f}_{\hat a}, \d \tilde{f}_{\hat b} ) = \frac{2}{\hbar} (1-x^0) (x^0 \, a_k b^k + a_0 \, b_k x^k)
	- \frac{2}{\hbar} \, a_0 x^0 \, b_k x^k -  \frac{2}{\hbar} \, a_k x^k \, b_l x^l \, ,
	\eeq 
which does not vanish on $\T_2^1$, meaning that the Jordan product $\langle \langle \tilde{f}_{\hat a}, \tilde{f}_{\hat b} \rangle \rangle$ defined in \eqref{Lie-Jordan-realizations} and \eqref{Pois-Bivector}, is not an element of the ideal $\mathcal{I}$.
To amend this problem we must modify $\tilde{G}$ to obtain a symmetrical bivector field whose associated product makes the affine closed subspace $\mathcal{I}$ an ideal of $\mathcal{F}(\D_2^\ast)$. However, in doing so, we must renounce to have a Jordan product realized in the space of linear functions.
Thus, the modified symmetric bivector field is given by \cite{Ciaglia-2017}
	\beq  \label{Rtilde}
	\tilde{\mathcal{R}} = \tilde{G} - \frac{2}{\hbar} \tilde{\Delta} \otimes \tilde{\Delta} \, ,
	\eeq
where $\tilde{\Delta} = x^\mu \frac{\partial}{\partial x^\mu}$ is the Euler--Liouville vector field. 
Then, it can be verified that
	\beq 
	\tilde{\mathcal{R}}(\d \tilde{f}_{\hat a}, \d \tilde{f}_{\hat b} ) 
	= \frac{2}{\hbar} \left(1-x^0 \right) \left(x^0 \, \delta^{kl} a_k b_l + (1 - 2 x^0) a^0 \, b_k x^k - 2 \, a_k x^k \, b_l x^l 		\right) 
	\in \mathcal{I} \, .
	\eeq
Hence, taking into account the pullback form the immersion map $i$, it is straightforward to obtain that the reduced symmetric tensor field $\mathcal{R}$ is
	\beq
	\mathcal{R}(\d f_{\hat{\sigma}^k}, \d f_{\hat{\sigma}^l}) := 
	f_{\hat{\sigma}^k \odot \hat{\sigma}^l} -  \frac{2}{\hbar} \, f_{\hat{\sigma}^k} \, f_{\hat{\sigma}^l} \, ,
	\eeq
where it can be seen that $\mathcal{R}$ does not leads to a Jordan product realization in the space of linea functions $\mathcal{F}(\T_2^1)$.
The expression of this symmetric bivector field in Cartesian coordinates is
	\beq \label{Symmetrical-Bivector}
	\mathcal{R} 
	= \frac{2}{\hbar} \left( \delta^{kl} - x^k x^l \right) \frac{\partial}{\partial x^k} \otimes \frac{\partial}{\partial x^l} \, .
	\eeq
Now that the vector space and the algebraic structure are compatible, the gradient-like vector field $Y_b$ associated with the linear function $f_{\hat b} = b_0 + b_k x^k$ is defined as
	\beq \label{Grad-like-vector-field}
	Y_b := \mathcal{R}( \d f_{\hat b} , \, \cdot \, ) 
	= \frac{2}{\hbar} \left( \delta^{kl}  - x^k x^l \right) b_k \frac{\partial}{\partial x^l} \ .
	\eeq
Note that this gradient-like vector field is quadratic in the Cartesian coordinate system adapted to $\T_2^1$.

To analize the behavior of the gradient-like vector field, one may compute the Lie derivative $\pounds$ of $r$ along the direction of the vector field  $Y_b$, i.e. $\pounds_{Y_b} r = Y_b(r)$, to obtain that 
	\beq
	\pounds_{Y_b} r = \frac{2}{\hbar \, r} \, \left( 1 -  r^2 \right) b_k \, x^k  \, ,
	\eeq
Then, the Lie derivative is different from zero if and only if $r \neq 1$; consequently, the gradient-like vector field is transversal to the leaves $\ell_r$ for $r\neq 1$ and only is tangent to the unit sphere which is the boundary of $\mathbf{B}$.
This fact allows us to observe that $Y_b$ does not change the rank of the density matrix at the boundary, therefore, a pure state remains pure under its evolution.


\subsection{ GKLS evolution on one q-bit systems} \label{subsection.2.3}

As we have seen in the previous subsection, for pure states the Hamiltonian and gradient-like vector fields give rise to a dynamical evolution that preserves the purity of the density matrix, that is, given initial conditions on the boundary of $\mathbf{B}$ the integral curves of $X_H$ and $Y_b$ remain on it. Consequently, if the system is prepared in a pure state, the Hamiltonian and gradient-like evolution take it to another pure state.
Now, we want to introduce a vector field that not only is transversal to the isospectral manifolds but also with the possibility of changing the rank of the density matrix for pure states.
For the finite-dimensional case, it is known that the GKLS generator $\mathbf{L}$ yields a quantum dynamical evolution described by linear equations~\cite{Gorini-1976, Lindblad-1976}. The GKLS master equation has the general form
	\beq \label{L-generator}
	\mathbf{L}(\xi) =  - \frac{\i}{\hbar} [\hat{H}, \hat{\xi} ]_{-} 
	- \frac{1}{2} \sum^3_j [ \hat{v}^\dagger_j \, \hat{v}_j , \hat{\xi} ]_{+}
	+ \sum^3_j \hat{v}_j \, \hat{\xi} \, \hat{v}^\dagger_j \, ,
	\eeq
 which may be expressed in terms of the Lie and the Jordan products, see Eq. \eqref{Lie-Jordan-algebra}, as the linear operator
	\beq \label{GKLS-Generator}
	\mathbf{L}(\hat{\xi}) = - [[\hat{H}, \hat{\xi} ]] - \frac{\hbar}{2} \, \hat{V} \odot \hat{\xi} 
	+ \hat{\mathcal{K}} ( \hat{\xi} ) \, ,
	\eeq
where $\hat{H} \in \D_2$, $\hat{V} = \sum^3_j  \hat{v}^\dagger_j  \, \hat{v}_j$ with  $\hat{v}_j \in \a_2$ and $\mathcal{K}(\hat{\xi})$ is a completely positive map given by
	\beq
	\hat{\mathcal{K}} (\hat{\xi}) = \sum^3_j \hat{v}_j \, \hat{\xi} \, \hat{v}^\dagger_j \, .
	\eeq
$\hat{\mathcal{K}} (\hat{\xi})$ is also called \emph{Choi--Kraus map}~\cite{Kraus-1971, Choi-1972, Choi-1975}.

We can define the vector field associated to the GKLS generator $\mathbf{L}$ using the following result \cite{Ercolessi-2010, Carinena-2015}: consider the linear map $A: \D^\ast_2 \to \D^\ast_2$ given by 
	\beq \label{linear-mapping-A}
	A(\xi) = A^\mu_{\ \nu} \, \xi^\nu \, e_\mu \, ,
	\eeq
where $A^\mu_{\ \nu}$ is the matrix representing the linear transformation and $\{ e_\mu \}$ denotes an orthonomal basis for $\D^\ast_2$. Thus, let $\{ x^\mu \}$ be the Cartesian coordinates associated to the orthonormal basis $\{ e^\mu \}$ of $\D_2$, it is possible to associate a vector field $Z_A \in \D_2^\ast$ (where $\D_2^\ast \cong$ to a cross-section of $T\D_2^\ast$) as
	\beq \label{linear-vector-field}
	Z_{A} := A^\mu_{\ \nu}\ x ^\nu \frac{\partial}{\partial x^\mu}.
	\eeq

Moreover, from this definition is direct to check that given $\hat{A}$ and $\hat{B}$ linear maps then $Z_{{A} + {B}} = Z_{A} + Z_{B}$ and $Z_{c A} = c \, Z_{A}$, with $c \in \C$. 
Then, the vector field in $\D^\ast_2$ associated to the GKLS generator $\mathbf{L}$ is \cite{Ciaglia-2017}
	\beq \label{Gamma-tilde}
	\tilde{\Gamma} = \tilde{X}_{H} - \frac{\hbar}{2} \tilde{Y}_{V} + \tilde{Z}_{\mathcal{K}},
	\eeq
where $\tilde{X}_{H}$ is the Hamiltonian vector field associated to $\hat{H}$ given by $\tilde{X}_{H} = \tilde{\Lambda}(\d f_{\hat{H}}, \cdot)$; $\tilde{Y}_{V}$ is the gradient-like vector field associated with $\hat{V}$ by means of $\tilde{Y}_{V} = \tilde{G}(\d f_{\hat{V}}, \cdot)$ and $\tilde{Z}_{\mathcal{K}}$ is the linear vector field associated with the completely positive map $\hat{\mathcal{K}}$. 

To find $\tilde{\Gamma}$ we start expressing the operators in the linear mapping \eqref{GKLS-Generator} in the form of \eqref{linear-mapping-A}. 
Thus, given the Hamiltonian operator $\hat{H} = H_\mu \, \hat{\sigma}^\mu$ and the density operator $\hat{\xi} = \frac{1}{2} \, x_\nu \, \hat{\sigma}^\nu$ we find that the products $[[\hat{H}, \hat{\xi}]]$  can be expressed as 
	\beq  \label{GKLS-operator-1}
	[[\hat{H}, \hat{\xi} ]] = \frac{1}{2} \, c^{\mu\nu}_{\eta} H_\mu \, x_\nu \, \hat{\sigma}^\eta 
	\eeq
and
	\beq
	\hat{V} \odot \hat{\xi} = \frac{1}{2} \, d^{\mu \nu}_{\eta} V_{\mu} \, x_\nu \, \hat{\sigma}^\eta \, .
	\eeq
Once defined these linear maps, we may associate the linea vector field on $\D^\ast_2$ from the definition \eqref{linear-vector-field}. 
Then, taking into account that $x^\mu = \delta^{\mu \nu} x_\nu$ and $\sigma_\mu = \delta_{\mu \nu} \hat{\sigma}^\nu$ we obtain
	\beq
	Z_{[[ {H}, {\xi}]]} 
	=  \delta_{\nu \sigma} \, \delta^{\eta \lambda} \, c^{\mu\nu}_{\eta}  H_\mu \, x^\sigma \frac{\partial}{\partial x^\nu}
	= - \, c_\sigma^{\lambda \nu} H_\lambda \, x^\sigma \frac{\partial}{\partial x^\nu}
	\eeq
and
	\beq
	Z_{V \odot \xi } 
	=  \delta_{\nu \sigma} \, \delta^{\eta \lambda} \, d^{\mu \nu}_{\eta}  
	V_\mu \, x^\sigma \, \frac{\partial}{\partial x^\nu}
	= d_\sigma^{\lambda \nu} \, V_\lambda \, x^\sigma \, \frac{\partial}{\partial x^\nu} \, .
	\eeq

On the other hand, to compute the completely positive map $\hat{\mathcal{K}}(\hat{\xi})$, let us note that it may be expressed in the form
	\beq
	\hat{\mathcal{K}}( \hat{\xi} ) = \frac{1}{2} \, \mathcal{K}_\mu \, \hat{\sigma}^{\mu}
	= \frac{1}{2} \, \Tr\{ \hat{\mathcal{K}}( \hat{\xi} ) \, \hat{\sigma}^\nu \} \delta_{\nu \mu} \, \hat{\sigma}^\mu \, ,
	\eeq
and by straightforward calculation we obtain
	\beq
	\Tr\{ \hat{\mathcal{K}}( \hat{\xi} ) \, \hat{\sigma}^0 \} 
	=  \sum^3_j \Tr\{ \hat{v}_j \, \hat{\xi} \, \hat{v}^\dagger_j \, \hat{\sigma}^0 \} 
	 = \Tr \left\{ \sum^3_j \hat{v}^\dagger_j \, \hat{v}_j \, \hat{\xi} \right\}
	 = \tilde{f}_{\hat{V}} \, ,
	\eeq
with $\tilde{f}_{\hat{V}} =  \Tr\{ \hat{\xi} \, \hat{V} \}$. 
Now, for the remaining components we have that
	\beq
	\Tr\{ \hat{\mathcal{K}}( \hat{\xi} ) \, \hat{\sigma}^k \} =
	\sum^3_j \Tr\{ \hat{v}_j \, \hat{\xi} \, \hat{v}^\dagger_j  \, \hat{\sigma}^k \} 
	= \frac{1}{2} \Tr\{ \hat{\mathcal{K}}(\hat{\sigma}^\eta)  \, \hat{\sigma}^k  \} x_\eta 
	= \frac{1}{2} \, \mathcal{K}^{\eta k} \, x_\eta \, .
	\eeq
where we have considered the definition $ \mathcal{K}^{\eta k} = \Tr\{ \hat{\mathcal{K}}(\hat{\sigma}^\eta)\,\hat{\sigma}^k \}$.
Consequently, the completely positive map $\hat{\mathcal{K}}( \hat{\xi} )$ has the form
	\beq
	\hat{\mathcal{K}}( \hat{\xi} ) =  \frac{1}{2} \tilde{f}_{\hat{V}} \, \hat{\sigma}^0 
	+ \frac{1}{4} \, \mathcal{K}^{\eta k} \, x_\eta \, \delta_{k m} \hat{\sigma}^m
	= \frac{1}{2} \tilde{f}_{\hat{V}} \, \hat{\sigma}^0 
	+ \frac{1}{4} \, \mathcal{K}^{\eta}_{\ m} \, x_\eta \, \hat{\sigma}^m \, .
	\eeq
Given this linear map we can now proceed to obtain the associated vector field
	\beq \label{jump-term}
	\tilde{Z}_{\mathcal{K}} = \tilde{f}_V \, \frac{\partial }{\partial x_0} 
	+ \frac{1}{2} \, \mathcal{K}^{\eta}_{\ m} \, \delta_{\eta \mu} x_\mu \, \delta^{m k}  \frac{\partial }{\partial x^k}
	= \tilde{f}_V \, \frac{\partial }{\partial x_0} 
	+ \frac{1}{2} \mathcal{K}_\mu^{\ k} \, x^\mu \, \frac{\partial }{\partial x^k} \, ,
	\eeq
then, the GKLS generator map $\mathbf{L}$ has associated the following linear vector field
	\beq \label{vector-field-tilde}
	\tilde{\Gamma} = - Z_{[[H, \xi ]]} - \frac{\hbar}{2} \, Z_{V \odot \xi } 
	+ \tilde{Z}_{\mathcal{K}} \, .
	\eeq
Comparing with the coordinate expression of the vector field in Eq. \eqref{Vector-fields} we observe that $Z_{[[H , \xi ]]} = - \tilde{X}_H$ and $Z_{V \odot \xi }  = \tilde{Y}_V$. Then, we have obtained an expression in terms of the coordinate basis for the vector field $\tilde{\Gamma}$ in Eq. \eqref{Gamma-tilde} associated to the GKLS master equation.	

As was done in the previous subsection, we need to find the vector field whose integral curves lies entirely in the affine space $\T_2^1$. To do so, we only need to apply the reduction procedure to $\tilde{\Gamma}$ by taking into account the immersion $i:= \T_2^1 \to \D_2^\ast$. Before performing the reduction, let us first express $\tilde{\Gamma}$ as follows
	\beq
	\tilde{\Gamma} 
	= \tilde{X}_{H} - \frac{\hbar}{2}\left( \tilde{Y}_{V} - \frac{2}{\hbar} \, \tilde{f}_{\hat{V}} \tilde{\Delta} \right) 
	+ \tilde{Z}_{\mathcal{K}}  - \tilde{f}_{\hat{V}} \tilde{\Delta} \, .
 	\eeq
Notice that the last term on the right hand side of the equation above is given in terms of the Cartesian coordinates as  
	\beq \label{Vector-Z}
	\tilde{Z}_{\mathcal{K}} - \tilde{f}_{\hat{V}} \tilde{\Delta} 
	= \left( 1 - x^0 \right) \tilde{f}_{\hat{V}} \frac{\partial }{\partial x^0} 
	+ \frac{1}{2} \mathcal{K}^k_{\ \mu} \, x^\mu \, \frac{\partial }{\partial x^k} 
	- \tilde{f}_{\hat{V}} \, x^k \frac{\partial}{\partial x^k} \, .
	\eeq
Now, the restriction of the vector field $\tilde{\Gamma}$ to the affine space to obtain $\Gamma$ can be easily obtained just setting $x_0 =1$. Thus, the first term in the right-hand-side of \eqref{Vector-Z} vanishes in $\T_2^1$. 
Therefore, $\tilde{Z}_{\hat{\mathcal{K}}} - \tilde{f}_{\hat{V}} \tilde{\Delta}$ by restriction gives 
	\beq \label{Choi-Kraus-VF}
	Z_{\mathcal{K}} = \frac{1}{2} \mathcal{K}^k_{\ \mu} \, x^\mu \, \frac{\partial }{\partial x^k} 
	- f_{\hat{V}} \, x^k \frac{\partial}{\partial x^k} 
	= \frac{1}{2} \mathcal{K}^k_{\ \mu} \, x^\mu \, \frac{\partial }{\partial x^k} - f_{\hat{V}}\, x^k \frac{\partial}{\partial x^k}\, ,
	\eeq
which will be called the Choi--Kraus vector field.
On the other hand, the vector field $ \tilde{X}_{\hat{H}}$ by restriction gives $X_{H}$ and $\tilde{Y}_{V} - \frac{2}{\hbar} \, \tilde{f}_{V} \tilde{\Delta}$ gives $Y_{V}$.
Then, upon the reduction we have that the quantum dynamical evolution generated by the GKLS generator $\mathbf{L}$ is described defining the GKLS vector field 
	\beq
	\Gamma = X_{H} + Y_{b} + Z_{\mathcal{K}} \\
	\eeq
in $\T_2^1$ by setting $\hat{b} = -  \frac{\hbar}{2} \hat{V}$.
Considering the Hamiltonian operator $\hat{H} = H_\mu \hat{\sigma}^\mu$ and $\hat{V} = V_\mu \hat{\sigma}^\mu$, then the GKLS vector field in Cartesian coordinates takes the following form
	\beq \label{GKLS-Cartesian-Coord}
	\Gamma = - \frac{2}{\hbar} \, \epsilon^{kj}_l H_k \, x^l \frac{\partial}{\partial x^j}
	-  \delta^{k j} \, V_k \, \frac{\partial}{\partial x^j}
	+ \frac{1}{2} \mathcal{K}^j_{\ \mu} \, x^\mu \, \frac{\partial }{\partial x^j} 
	- V_0\, x^k \frac{\partial}{\partial x^k} \, .
	\eeq
It is interesting that the nonlinear term in $Z_{\mathcal{K}}$ and $Y_{V}$ cancel out in the sum $- \frac{\hbar}{2} \, Y_{V} + Z_{\mathcal{K}}$ and then, the vector field $\Gamma$ for a $q$-bit system in Cartesian coordinates is linear.

Therefore, the GKLS dynamics allows for a \emph{decomposition principle}, i.e. the conservative part is given by the Hamiltonian part as a reference dynamics, while the sum of the gradient-like and the Choi--Kraus vector fields can be considered as a ``perturbation term'' associated with dissipation.
	

\subsection{Damping phenomena in two-level atomic system} \label{subsec.2.4}

In this subsection we analyze two simple cases to illustrate the use of the GKLS vector field to determine the dynamics of a physical system. Let us consider a two-level atom with ground state $| 1 \rangle$ and excited state $| 2 \rangle$, i.e., we are considering eigenstates of the Hamiltonian $\hat{H}_{\tiny \mbox{atm}}$ with eigenvalue $E_1$ and $E_2$, respectively. 
This allows to define the transition operators
	\beq
	\hat{\sigma}_{i j} = | i \rangle \langle j | \, ,
	\eeq
with $i, j = 1,2$.
Then, $\hat{H}_{\tiny \mbox{atm}}$ may be written as
	\beq \label{Atm-Hamiltonian}
	\hat{H}_{\tiny \mbox{atm}} = \sum_i E_i\, | i \rangle \langle i |
	= \sum_i E_i\, \hat{\sigma}_{ii} = \frac{1}{2}(E_1 + E_2) \hat{\sigma}^0 + \frac{1}{2}(E_1 - E_2) \hat{\sigma}^3 \, .
	\eeq  
Now, defining $\hbar \, \nu = E_1 - E_2$ and for simplicity ignoring constant terms, the atomic Hamiltonian takes the form
	\beq
	\hat{H}_{\tiny \mbox{atm}} =  \frac{1}{2} \, h \, \nu \, \hat{\sigma}^3 \, ,
	\eeq 
with $h \nu$ the transition energy between the states. 
Furthermore, taking into account the Pauli matrices we can express the transition operators as
	\beq
	\hat{\sigma}_{12} = \hat{\sigma}_+ = \frac{1}{2}(\hat{\sigma}^1 + \i \, \hat{\sigma}^2 )
	\quad
	\text{and}
	\quad
	\hat{\sigma}_{21} = \hat{\sigma}_- = \frac{1}{2}(\hat{\sigma}^1 - \i \, \hat{\sigma}^2 )\, , 
	\eeq
which represent the transition between states, i.e., $\hat{\sigma}_{21} = \hat{\sigma}_- $ represents the transition $|1 \rangle \to |2 \rangle$ and $\hat{\sigma}_{12} = \hat{\sigma}_+$ the transition $|2 \rangle \to |1 \rangle$.  
Then, associated to the Hamiltonian $\hat{H}_{\tiny \mbox{atm}}$ we have the vector field
	\beq
	X_{\hat{H}_{\tiny \mbox{atm}}}   = - \nu \, \left( x^2 \frac{\partial}{\partial x^1} - x^1 \frac{\partial}{\partial x^2} \right)\, .
	\eeq
	
	\begin{figure}[! h t]
	\centering
	\includegraphics[width = 7.5 cm]{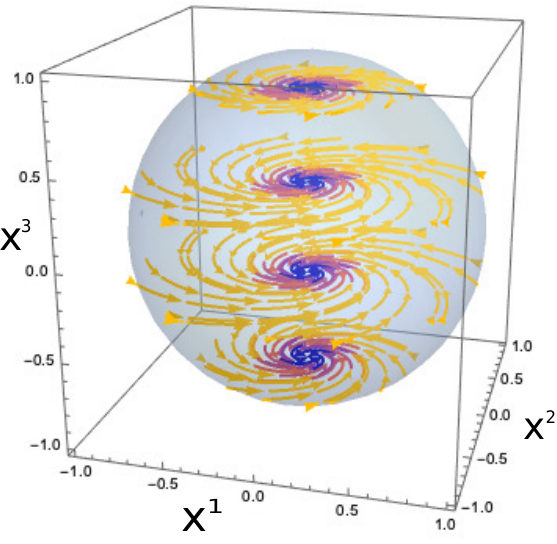}
	\caption{In these figures we have considered GKLS vector field for a 2-level atom system 
	$\hat{H}_{\tiny \mbox{atm}} =  \frac{1}{2} \, h \, \nu \, \sigma^3$ with a dissipative term introduced by means of the 		operator $\hat{v}_1 = \sqrt{\frac{\gamma}{2}} \, \hat{\sigma}^3$.
	The GKLS vector field has been displayed in Cartesian coordinates $(x^1,x^2,x^3)$, where we have considered 
	the frequency $\nu = 2$ and the damping parameter $\gamma = 1$.}
	\label{Fig-2}
	\end{figure}
	
Let us now introduce a dissipative evolution by means of the GKLS formalism by considering the simplest case corresponding to $\hat{v}_1 = \sqrt{\frac{\gamma}{2}} \, \hat{\sigma}^3$, where $\gamma$ is a constant with dimensions of frequency that modulates the dissipation, called \emph{the damping parameter}. In this case, we have that $\hat{V} = \hat{v}^\dagger_1 \, \hat{v}_1 = \gamma \, \hat{\sigma}^0$, thus $Y_{\hat{V}} = 0$; therefore, the dissipation to the Hamiltonian system is exclusively determined by the Choi--Krauss vector field. 

To compute the vector field $Z_\mathcal{K} $ in Eq. \eqref{Choi-Kraus-VF}, we first note that $f_{\hat{V}} = \frac{\gamma}{2}$ and $ \mathcal{K}^k{}_0 = 0$, then
	\beq
	\frac{1}{2} \mathcal{K}^k{}_j \, x^j \, \frac{\partial }{\partial x^k} 
	= \frac{1}{2} \Tr\{ \hat{\mathcal{K}}(\sigma_j)  \, \hat{\sigma}^k \} x^j \, \frac{\partial }{\partial x^k}
	=  \frac{\gamma}{4} \delta_{j l} 
	\Tr\{ \hat{\sigma}^3 \, \hat{\sigma}^l \, \hat{\sigma}^3 \, \hat{\sigma}^k \} x^j \, \frac{\partial }{\partial x^k} \, .
	\eeq
Taking into account the property $\Tr\{ \hat{\sigma}^i \, \hat{\sigma}^j \, \hat{\sigma}^k \, \hat{\sigma}^l \} = 2( \delta^{i j} \delta^{k l} - \delta^{i k} \delta^{j l} + \delta^{i l} \delta^{j k} )$ it is direct to show that 
	\beq
	\frac{1}{2} \mathcal{K}^k{}_j \, x^j \, \frac{\partial }{\partial x^k} 
	= \gamma \, x^3 \frac{\partial }{\partial x^3} - \frac{\gamma}{2} \, x^k \frac{\partial}{\partial x^k}\, .
	\eeq	
Then, the Choi--Krauss vector field in Cartesian coordinates takes the following form 
	\beq
	Z_{\mathcal{K}} = - \gamma \left(
	x^1 \frac{\partial}{\partial x^1} + x^2 \frac{\partial}{\partial x^2}
	\right) \, .
	\eeq
Once calculated the Hamiltonian and Choi--Kraus terms, we can finally give the expression for the GKLS vector field 
	\beq\label{decoherence-vec-field}
	\Gamma = - ( \nu \, x^2 + \gamma \, x^1) \frac{\partial}{\partial x^1} 
	+ ( \nu \, x^1 - \gamma \, x^2) \frac{\partial}{\partial x^2} \, .
	\eeq	
The vector field $\Gamma$ is displayed in Fig. \ref{Fig-2}, where the ball $\mathbf{B}$ is also plotted considering the Cartesian coordinates $(x^1, x^2, x^3)$.
We observe that the line $(0, 0, x^3)$ is singular and behaves as an attractor and then any pure state converge as a state with some statistical mix. 
In particular for the initial condition with $x^3 = 0$ the vector field leads to the state with a maximal degree of mixture.
\\
\\

\begin{figure}[! h t]
	\centering
	\includegraphics[width = 7.5 cm]{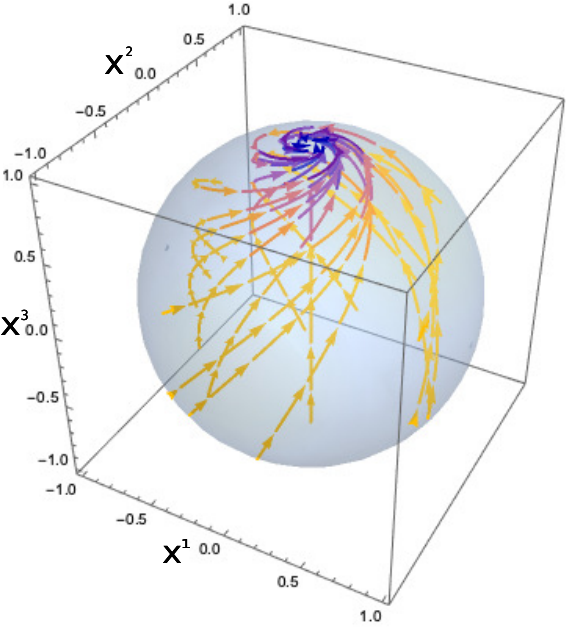}
	\caption{In this figure we plot the GKLS vector field for a 2-level atom system 
	$\hat{H}_{\tiny \mbox{atm}} =  \frac{1}{2} \, h \, \nu \, \hat{\sigma}^3$ with dissipative part introduced by means of the 	operator $\hat{v}_1= \sqrt{2 \,\gamma} \, \hat{\sigma}_{+}$. Here we have considered the Cartesian coordinates 
	$(x^1, x^2, x^3)$ in the ball, with parameters $\nu = 2$ and $\gamma = 1$. }
	\label{Fig-3}
	\end{figure}

Another interesting case is to consider the transition operators to model a different dissipative dynamics. 
Let $v_1~=~ \sqrt{2 \,\gamma} \, \sigma_{+}$, $\hat{V} = \hat{v}^\dagger_1 \, \hat{v}_1 = \gamma \, (\hat{\sigma}^0 - \hat{\sigma}^3)$ and $f_{\hat{V}} = \gamma \, (1 - x^3)$. By a straightforward calculation we find that
	\beq
	-  \delta^{k j} \, V_k \, \frac{\partial}{\partial x^j} = \gamma \, \frac{\partial}{\partial x^3} \, ,
	\qquad
	- V_0 \, \Delta = - \gamma \, x^k \frac{\partial}{\partial x^k}
	\eeq
and 
	\beq 
	\frac{1}{2} \mathcal{K}^k{}_\mu \, x^\mu \, \frac{\partial }{\partial x^k} =
	 \frac{1}{2} \Tr\{ \hat{\mathcal{K}}(\sigma_0)  \, \hat{\sigma}^k \} \, \frac{\partial }{\partial x^k} +
	\frac{1}{2} \Tr\{ \hat{\mathcal{K}}(\sigma_j)  \, \hat{\sigma}^k \} x^j \, \frac{\partial }{\partial x^k}
	= \gamma \, (1 - x^3) \frac{\partial}{\partial x^3} \, .
	\eeq
Then, the GKLS vector field is given by
	\beq \label{Gamma2}
	\Gamma = - \left( \nu \, x^2 + \gamma \, x^1 \right) \frac{\partial}{\partial x^1} 
	+ \left(\nu \, x^1 - \gamma \, x^2 \right) \frac{\partial}{\partial x^2} 
	+2 \, \gamma \, (1 - x^3) \frac{\partial}{\partial x^3} \, .
	\eeq	
We see that the components in the $x^1$ and $x^2$ directions are the same as those of the previous case, c.f. \eqref{decoherence-vec-field}; nevertheless, in the present case $\Gamma$ has a component in the $x^3$ direction. The integral curves corresponding to \eqref{Gamma2} are the solutions to the {\it linear} system of equations  
	\begin{align}
	\dot{x}^1 & = - \nu \, x^2 - \gamma \, x^1 \, , \nonumber \\
	\dot{x}^2 & = \nu \, x^1 - \gamma \, x^2 \,  ,  \nonumber \\
	\dot{x}^3 & =  2 \, \gamma \, (1 - x^3) \, .
	\end{align}
From this vector field we see that there is only one singular point in $(0,0,1)$, i.e., the ``north pole'' of the sphere, which is an attractor. 
The behaviour of this vector field is illustrated in Fig. \eqref{Fig-3}.  We notice that regardless of the initial condition the evolution converges to the ``north pole'' of the sphere.
		

\section{GKLS dynamics on Gaussian states} \label{sec.3}

In this section, we are interested in obtaining the GKLS vector field for a class of systems whose quantum states are described by Gaussian states. We will follow, in general, the same steps taken for constructing the vector field for the q-bit systems and we will apply the result to describe a quantum harmonic oscillator with dissipation. 
Before we begin the procedure to construct the GKLS vector field, let us first review some important properties of Gaussian states.

A Gaussian state in the position space representation has the general form
	\beq \label{Gaussian-Density-Matrix}
	\langle q' | \, \hat{\rho} \, | q \rangle 
	=  \frac{1}{\sqrt{2\, \pi \, \sigma^2_{q}}}  
	\exp \Bigg\{
	 \frac{1}{2 \, \sigma^2_{q} }
	\left[
	\frac{\sigma_{qp}}{\hbar} (q - q' ) - \frac{\i}{2} \, ( q + q' - 2 \langle \hat{q} \rangle )
	\right]^2 - \frac{ \sigma^2_{p} }{ 2 \, \hbar^2} (q - q')^2  + \frac{\i}{\hbar} \, \langle \hat{p} \rangle \, (q - q') 				\Bigg\} \, ,
	\eeq
where $q$ and $q'$ are the coordinates in the position space for two different points. The uncertainty for each operator is defined as 
	\beq \label{uncertainties}
	\sigma^2_{q} =  \langle \hat{q}^2 \rangle - \langle \hat{q} \rangle^2\,\qquad \sigma^2_{p} = \langle \hat{p}^2 \rangle - \langle \hat{p} \rangle^2 \, , 
	\eeq 
and the correlation between the position and momentum operators is
	\beq \label{correlation}
	\sigma_{qp} 
	= \frac{1}{2} \langle \hat{q} \, \hat{p} + \hat{p} \, \hat{q} \rangle - \langle \hat{q} \rangle \langle \hat{p} \rangle \, .  
	\eeq
For simplicity, in the following we will consider that the expectation values of position and momentum are zero, i.e. $(\langle \hat{q} \rangle , \langle \hat{p} \rangle) = (0,0)$.

A simplified expression of the Gaussian state can be obtained in the Wigner representation $W_r (q,p)$ of $\hat{\rho}$ by applying the \emph{Wigner-Weyl transformation}~\cite{Weyl-1927} to obtain the Wigner quasi-distribution function\footnote{The Wigner-Weyl transform of $\hat{\rho}$ is given by 
	$$W(q,p) = \frac{1}{2\pi \hbar} 
	\int_{-\infty}^{\infty} e^{-ipy/\hbar} \langle q + y/2|\hat{\rho}|q - y/2\rangle dy \, . $$ }	 
	\beq \label{Wigner-function}
	W_r( q , p ) = \frac{1}{\pi \, \hbar \, r} \exp \left\{
	- \frac{2}{\hbar^2 \, r^2} \left[
	\sigma_{p}^2 q^2 
	- 2 \sigma_{qp} q p
	+ \sigma_{q}^2 p^2
	\right]
	\right\} \, ,
	\eeq
where $(q,p)$ denote points in phase space and the parameter $r \in [1 , \infty)$, related to the Robertson-Schr\"odringer uncertainty relation, is defined as
	\beq \label{SR-Ineq}
	\sigma_{q}^2 \, \sigma_{p}^2  -  \sigma^2_{qp} = \frac{\hbar^2 \, r^2}{4} \, .
	\eeq 
This relation identifies a coadjoint orbit of the group $SL(2,\mathbb{R})$ on its Lie algebra.	
In addition, it can be shown by direct calculation that the \emph{degree of non-purity of a Gaussian state} is given by the parameter $r$, due to
   	\beq \label{constrain-GS}
	\Tr\{ \hat{\rho}^2 \} = \frac{1}{r} \, .
	\eeq
In particular, when $r=1$ the density Gaussian matrix or the Wigner function corresponds to the \emph{generalized coherent states}, see for instance Ref.~\cite{Cruz-2021, Perelomov-1977,  Zhang-1990, Cruz-2015}. 
Hence, in this case, the Gaussian density matrix may be factorized as $\hat{\rho} = | 0 \rangle \langle 0 |$ where $| 0 \rangle$ denotes the vacuum state of the Fock states.
	

The normalized positive functional  $\langle q' | \, \hat{\rho} \, | q \rangle$ or equivalently its associated Wigner function $W_r( q , p )$ describing the quantum states  is parametrized by $( \sigma_{q}^2,  \sigma_{p}^2,  \sigma_{qp} )$ and these parameters are constrained by the relation \eqref{SR-Ineq}. 
We are interested in defining a manifold for the entire space of states parametrized by the uncertainties and the correlation.
In order to do so, we consider the immersion of a finite-dimensional manifold into the space of normalized positive functionals $\mathcal{L}^1$ by means of a \emph{Weyl map}~\cite{Ercolessi-2010}. 

\begin{figure}[! t]
	\centering
	\includegraphics[width = 12.5 cm]{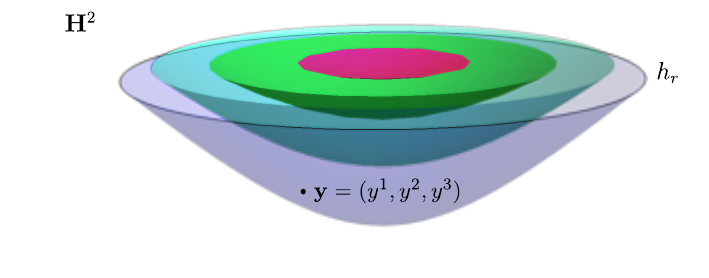}
	\caption{Pictorial representation of the space of the solid $\mathbf{H}$. 
	A Gaussian state is represented by a point $\mathbf{y} = (y^1, y^2, y^3)$ in $\mathbf{H}$.
	In addition, one may see that this solid is foliated by a the disjoint family of hyperboloids $h_r$. }
	\label{Fig-4}
	\end{figure}

To introduce the Weyl map, it is important to first make some remarks of the space of parameters.
To describe the immersion of the space of second momenta $( \sigma_{q}^2,  \sigma_{p}^2,  \sigma_{qp} )$, one intoduces the following set of coordinates  
	\beq \label{hr-to-Sr}
	y^1 =  \frac{1}{\hbar} (\sigma^2_{p} + \sigma^2_{q}) \, ,
	\qquad
	y^2 =  \frac{2}{\hbar} \, \sigma_{qp} \, , 
	\qquad
	y^3 =  \frac{1}{\hbar} (\sigma^2_{p} - \sigma^2_{q}),
	\eeq
such that the constraint \eqref{SR-Ineq} defines the manifold for a fixed value of $r$ as
	\beq \label{Foliation-H}
	h_r = \{ (y^1, y^2, y^3) \in \R^3 \, | \, (y^1)^2 - (y^2)^2 - (y^3)^2 = r^2 \} \, ,
	\eeq
where $y^1 > 0$. 
The manifolds $h_r$ are consequently upper-hyperboloids in $\R^3$ known as \emph{pseudo-spheres}, in analogy to the spheres in the q-bit case \cite{Balazs-1986}. 
From the expression \eqref{constrain-GS}, it is clear that each manifold $h_r$ has a different degree of statistical mixture, while the purity condition $\Tr\{ \hat{\rho}^2\} = 1$ identifies the hyperboloid
	\beq 
	H^2 = \{ (y^1, y^2, y^3) \in \R^3 \, | \, (y^1)^2 - (y^2)^2 - (y^3)^2 = 1 \} \, .
	\eeq
On the other hand, the condition $\Tr\{\hat{\rho}^2\} < 1$ identifies the non-pure states as the set defined by
	\beq
	(y^1)^2 - (y^2)^2 - (y^3)^2 > 1 \, ,
	\eeq
where the maximal ``impurity'' is obtained for the state in the limit $r \to \infty$. Therefore, a general Gaussian state is parametrized by points $\mathbf{y} = (y^1, y^2, y^3)$ in the three-dimensional differential manifold
	\beq \label{Sold-Hyperboloid}
	\mathbf{H} = \{ (y^1, y^2, y^3) \in \R^3 \, | \, (y^1)^2 - (y^2)^2 - (y^3)^2 \geq 1 \} \, .
	\eeq
The space $\mathbf{H}$ is a differential manifold with boundary and it can be described as the foliation given by the hyperboloids \eqref{Foliation-H} labeled by $r$ as
	\beq
	\mathbf{H} = \bigcup_{r \geq 1} h_r \, ,
	\eeq  
where the leaves of this foliation are the coadjoint orbits of $SL(2,\mathbb{R})$, and they are defined in Eq.~\eqref{Foliation-H} and on each leaf a differential structure can be given. 
A schematic picture of this foliation is displayed in Fig.~\ref{Fig-4}.

There is a clear analogy between the ball $\mathbf{B}$ employed in the description of q-bit systems and the solid hyperboloid $\mathbf{H}$, in the sense that they are manifolds containing the information of the quantum states and both manifolds are foliated such that each leaf of the foliation has a fixed degree of purity.  
While in the q-bit case, each orbit is associated with a given spectrum, here it is associated with purity.

We are interested in describing the immersion of each leaf $h_r$ into the space of normalized positive functionals $\mathcal{L}^1$, then, we will require to find a set of coordinates for $h_r$. For instance, one may introduce hyperbolic coordinates by considering the mapping $\nu:h_r \to \C$, where each point $(y^1, y^2, y^3)$ on $h_r$ is described by a point in the complex plane $\kappa_r = \frac{\tau}{2} \, \e^{\i \, \varphi}$ given by 
	\begin{align}\label{hyp-coor}
	y^1 = r \cosh{\tau}\, ,
	\quad
	y^2 = r \sinh{\tau} \cos{\varphi} \, , 
	\quad
	y^3 = r \sinh{\tau} \sin{\varphi} \, ,
	\end{align}
The coordinates $(\tau , \varphi )$ are called squeezing parameters in quantum optics~\cite{Scully-1999}.	

Now we are in position to introduce the Weyl map from $h_r$ to the set of unitary operators
	\beq
	\hat{S}(\xi_r) \equiv \exp\left[ \bar{\kappa}_r \, \hat{K}_{-} - \kappa_r \, \hat{K}_{+} \right]
	\eeq 
known as \emph{squeezing operators}, where the operators $\hat{K}_{\pm}$ together with $\hat{K}_0$ are the generators of the Lie algebra $\su(1,1)$ satisfying the commutation relations
	\beq
	[\hat{K}_{+}, \hat{K}_{-}] = - 2 \, \hat{K}_0 \, , 
	\quad
	[\hat{K}_0, \hat{K}_\pm] = \pm \, \hat{K}_{\pm} \, .
	\eeq
Therefore, we have defined the map $W : h_r \to \mathcal{U}(\h): \kappa_r \mapsto \hat{S}(\kappa_r)$,
denoting $\mathcal{U}(\h)$ the set of unitary operators on a Hilbert space $\h$, where the unitary representation of $h_r$ is given by the squeezing operators $\hat{S}(\kappa_r)$.
Furthermore, the Weyl map allows to introduce the immersion
	\beq \label{immersion}
	w : h_r \to \mathcal{L}^1 \, ,
	\eeq
defined in the following form, given a fiducial state $\rho_0 \in \mathcal{L}^1$ we define the action of the map $w$ as
	\beq
	w : \kappa_r \mapsto \hat{\rho}(\kappa_r) \equiv 
	\hat{S}(\kappa_r) \, \hat{\rho}_0 \, \hat{S}^\dagger(\kappa_r) \, .
	\eeq 
To be more specific, let us consider as a fiducial state the Gaussian state with zero squeezing parameters $(\tau_r, \varphi_r) = (0, 0)$, i.e., the state has zero correlation $\tilde{\sigma}_{qp} = 0$, therefore, the fiducial Gaussian state in the position representation is
	\beq \label{Non-Sque}
	\langle q' | \, \hat{\rho}_0 \, | q \rangle  
	=  \frac{1}{\sqrt{2\, \pi \, \tilde{\sigma}^2_{q} } }
	\exp \Bigg\{
	- \frac{1}{8 \, \tilde{\sigma}^2_{q}} ( q + q' )^2 
	- \frac{ \tilde{\sigma}^2_{p} }{ 2 \, \hbar^2} (q - q')^2  \Bigg\} \, .
	\eeq	
Acting with the squeezing operator there is a translation in $h_r$ from $(0, 0)$ to $(\tau_r, \varphi_r)$ which in terms of the uncertainties and the correlation is given by the map $( \tilde{\sigma}^2_{q}, \tilde{\sigma}^2_{p}, 0)  \mapsto (\sigma^2_{q}, \sigma^2_{p}, \sigma_{qp})$ defined by the transformation 
	\begin{align} \label{Translation-Sigma}
	\sigma^2_{q} & = \tilde{\sigma}^2_{q}
	\left[ \, \cosh{\tau_r} - \cos{\varphi_r} \, \sinh{\tau_r} \, \right] \, , \nonumber \\
	\sigma^2_{p} & = \tilde{\sigma}^2_{p}
	\left[ \, \cosh{\tau_r} + \cos{\varphi_r} \, \sinh{\tau_r} \, \right] \, ,  \nonumber \\
	\sigma_{qp} & = - \frac{ \hbar \, r}{2} \, \sin{\varphi_r} \, \sinh \tau_r \, ,
	\end{align}
obtaining the Gaussian density matrix 
	\beq
	\langle q' | \hat{\rho}(\kappa_r) | q \rangle 
	=  \frac{1}{\sqrt{2\, \pi \, \sigma^2_{q} }}  
	\exp \Bigg\{
	 \frac{1}{2 \, \sigma^2_{q} }
	\left[
	\frac{ \sigma_{qp} }{\hbar} (q - q' ) - \frac{\i}{2} \, ( q + q' ) \right]^2 
	- \frac{ \sigma^2_{p} }{ 2 \, \hbar^2} (q - q')^2  \Bigg\} \, .
	\eeq
Let us note that because of the highly non-linearity of the squeezing parameters $(\tau_r, \varphi_r)$, it will result convenient to employ a set of coordinates adapted to the upper half complex plane 
	\beq \label{Riccati-coordinates}
	\text{HP}_r = \left\{ \Ri \in \C \, | \, \Ri_{\tiny\mbox{I}} > 0 \right\} \, ,
	\eeq
where $\Ri = \Ri_{\tiny\mbox{R}} + \i \, \Ri_{\tiny\mbox{I}}$ and defined by the relations
	\beq  \label{C-coordinates}
	y^1 = \i \, r \, \frac{1 + \Ri \bar{\Ri}}{ \Ri - \bar{\Ri} }  \, ,
	\qquad
	y^2 = \i \, r \, \frac{ \Ri + \bar{\Ri} } { \Ri - \bar{\Ri} } \, ,
	\qquad
	y^3 = \i \, r \frac{ - 1+ \Ri \bar{\Ri} } { \Ri - \bar{\Ri} } \, .
	\eeq
The two-dimensional space for $r = 1$ described by these coordinates is known in the literature as the \emph{Siegel upper half plane}~\cite{Siegel-1943}.	 


Therefore, the immersion map \eqref{immersion} allows one to define a one-to-one relation between Gaussian states and points in the parameter space given by points in the solid $\mathbf{H}$.
Moreover, because a Gaussian initial state remains Gaussian under evolution obtained by operators that are at most quadratic in position and momentum~\cite{Bonet-Luz-2016}, we restrict our study to observables of the form
	\beq \label{H-Canonica}
	\hat{H} = \frac{1}{2} \left[ H_1 \, \hat{p}^2 + V(\hat{q} \, \hat{p} + \hat{p} \, \hat{q}) + H_2 \, \hat{q}^2 \right] \, ,
	\eeq
where $H_1$, $H_2$ and $V$ are real constants. 
The expectation value of this operator defined by $f_{\hat{H}} :=  \Tr \{ \hat{\rho} \, \hat{H} \} $, is given by 
	\beq
	f_{ \hat H} =
	 \frac{1}{2} \left[ H_1 \, \langle \hat{p}^2 \rangle+ V \langle \hat{q} \, \hat{p} + \hat{p} \, \hat{q}\rangle 
	+ H_2 \langle \hat{q}^2 \rangle \right]
	=  \frac{1}{2} \left[
	 H_1 \, \sigma^2_p + 2 \, V \sigma_{qp} + H_2 \sigma_q^2 \, ,
	 \right] \, 
	\eeq
i.e., the expectation value are linear functions of the second moments. 
In particular, if we consider the following set of operators
	\beq \label{SL-Lie-algebra}
	\hat{L}^1 = \frac{1}{4} \left( \hat{p}^2 + \hat{q}^2 \right) \, ,
	\quad
	\hat{L}^2 = \frac{1}{4} \left( \hat{q}\,\hat{p} + \hat{p}\,\hat{q} \right)
	\quad
	\text{and}
	\quad
	\hat{L}^3 = \frac{1}{4} \left( \hat{p}^2 - \hat{q}^2 \right) \, ,
	\eeq
any arbitrary operator in the form \eqref{H-Canonica} can be expressed as
	\beq
	\hat{H} = \tilde{H}_k \hat{L}^k \, ,
	\quad
	\text{with}
	\quad
	k = 1,2,3 \, ,
	\eeq
by setting
	\beq
	\tilde{H}_1 = H_1 + H_2,
	\quad
	\tilde{H}_2 = 2 V
	\quad 
	\text{and}
	\quad
	\tilde{H}_3 = H_1 - H_2 \, .
	\eeq	
Thus, taking into account that
	\beq
	\Tr\{ \hat{L}^k \hat{\rho} \}
	= \frac{\hbar}{4} y^k \, ,
	\eeq
the expectation value of any quadratic operator in the position and momentum operators is given by
	\beq
	f_{ \hat H} = \frac{\hbar}{4}  \tilde{H}_\mu y^\mu \, ,
	\eeq	
which are linear functions on the space of parameters $\mathbf{H}$.
Therefore, we restrict our study to consider only observables that are quadratic in the position and momentum operators because they not only preserve the Gaussian form of the state under its (Hamiltonian) evolution, but also they can be represented as linear functions on the space of parameters.
To conclude this section, it is important to note that the set of operators $\{ \hat{L}^k \}_{k=1,2,3}$ in Eq. \eqref{SL-Lie-algebra} closes on a Lie algebra 
given by the commutation relations
	\beq \label{sl-Lie-product}
	[[\hat{L}^1, \hat{L}^2]] = \hat{L}^3 \, , 
	\quad
	[[\hat{L}^2, \hat{L}^3]] = - \, \hat{L}^1
	\quad
	\text{and}
	\quad
	[[\hat{L}^1, \hat{L}^3]] = - \, \hat{L}^2 \, .
	\eeq
Additionally, there is a direct connection between the operadores $\{\hat{K}_{\pm}, \hat{K}_{0} \}$ of the $\su(1,1)$ Lie algebra and the operator $\{ \hat{L}^k \}_{k=1,2,3}$ because
	\beq
	\hat{K}_{\pm} = \frac{1}{\i \hbar} \left( \hat{L}^3 \pm \i \hat{L}^2 \right)
	\quad
	\text{and}
	\quad
	\hat{K_0} =  \frac{1}{\i \hbar}  \hat{L}^1 \, .
	\eeq 

As we have seen for the q-bit system, a fundamental property of the observables is the fact that they may be considered as elements of a finite $C^\ast$-algebra. 
In this case, we consider the following set of matrices to represent the operators
	\beq \label{SL-algebra-basis}
	\hat{L}^0 = \frac{\hbar}{2} \, \left(
	\begin{array}{c c}
	1 & 0  \\
	0 & 1   \\    
	\end{array}
	\right)\, ,
	\quad
	\hat{L}^1 = \frac{\i \, \hbar}{2} \,  \left(
	\begin{array}{c c}
	0  &  1  \\
	-1 &   0   \\    
	\end{array}
	\right) \, , 
	\quad
	\hat{L}^2 = \frac{\i \, \hbar}{2} \, \left(
	\begin{array}{c c}
	1  &  0  \\
	0 &   -1   \\    
	\end{array}
	\right) \, ,
	\quad
	\hat{L}^3 = \frac{\i \, \hbar}{2} \, \left(
	\begin{array}{c c}
	0  &  1  \\
	1 &   0   \\    
	\end{array}
	\right) \, ,
	\eeq
where the $\{\hat{L}^k\}$, with $k=1,2,3$, is a non-unitary matrix realization of the Lie algebra defined by the commutation relations in \eqref{sl-Lie-product}.
Moreover, the matrices $\{ \hat{L}^\mu \}$ obey the relations
	\beq \label{Jordan-Product-sl}
	 \hat{L}^\mu \odot \hat{L}^\nu = \, g^{\mu \nu} \, \hat{L}^0
	\eeq
where $\odot$ is the Jordan-product defined in Eq. \eqref{Lie-Jordan-algebra}, with $\hat{\mathbb I}$ the identity operator and $g^{jk}$ being the entries of the diagonal matrix
	\beq \label{metric-4}
	g =  \left(
	\begin{array}{c c c c}
	1 & 0 & 0 & 0 \\
	0 & 1 & 0 &  0  \\
	0 & 0 & -1&  0  \\
	0 & 0 & 0 & -1    
	\end{array}
	\right) \, .
	\eeq
Therefore, any observable may be expressed in term of the matrix representation as $\hat{a} = a_\mu \hat{L}^\mu$, with $\mu = 1,2,3,4$. Just as is the q-bit case, these operators are elements of a $C^\ast$-algebra, and besides, they are elements of a Lie--Jordan algebra.

If we denote the space of observables as $\D$, we may also introduce the dual space $\D^\ast$ such that for every element $\hat{a} \in \D$ and $\xi \in \D^\ast$ there is a linear function $f_{\hat{a}}$ on $\D^\ast$ 
	\beq
	f_{\hat{a}} := \langle \xi , \hat a \rangle
	\eeq
where the pairing between elements of the algebra and its dual is given by the expression $\langle \xi , \hat a \rangle = \Tr \{ \hat{\xi} \, \hat{a} \}$. 
Thus, in Cartesian coordinates the element $\xi \in \D^\ast$ has the form
	\beq \label{xi-func}
	\xi =  \frac{1}{2\hbar} \, y^\mu\, L_\mu \, ,
	\eeq
 where $L_\mu \equiv g_{\mu\nu} \hat{L}^\nu$, $y^\mu \equiv g^{\mu\nu}y_\nu$, and $g_{\mu\nu}$ are the components of the matrix \eqref{metric-4}.The components $\{\frac{\hbar}{4} y^\mu\}$ are defined by the following relation
 	\beq \label{coords-y}
	f_{\hat{L}^\mu} 
	= \langle \xi , \hat{L}^\mu \rangle 
	= \Tr\{ \hat{L}^\mu \, \hat{\xi}  \} = \frac{\hbar}{4} y^\mu \, .
	\eeq
Consequently, we have that the linear function $f_{\hat{a}}$, associated with the observable $\hat{a} = a_\mu \, \hat{L}^\mu \in \D$, corresponds to 
	\beq
	f_{\hat{a}} = \langle \xi , \hat{a} \rangle = \frac{\hbar}{4} a_\mu y^\mu \, ,
	\eeq
which coincides with the expectation value of the operator.
Therefore, the matrix representation of the operator allows one to follow a similar procedure for the q-bit case to obtain the GKLS vector field.

	
\subsection{Dynamical study of Gaussian state systems from a state point of view} \label{subsec.3.1}

In quantum mechanics, under the usual probabilistic interpretation, any transformation of a state is described by a one-parameter group of unitary transformations, in particular, the time evolution of a state, i.e., the probability conservation is secured by the Schr\"odinger equation. The assumption of this interpretation, then requires the infinitesimal generator to be a self-adjoint operator acting on the separable complex Hilbert space $\h$ associated with the physical system. Let us describe, for the case of Gaussian states, how this unitary evolution can be cast in terms of a Hamiltonian vector field. To do so, we will make use of the K\"ahler structure that the upper-half-hyperboloid $h_r$ bears, which later will probe to be useful when we extend the dynamics adding dissipation as a perturbation term.

As we have seen in the previous subsection, the space $\mathbf{H}$ of all the quantum states described by Gaussian density matrix is foliated by leaves labeled by $r$. Each leaf $h_r$ is endowed with a K\"ahler structure $(\omega_{h_r}, g_{h_r}, J_{h_r})$, then it is possible to introduce symplectic and gradient dynamics on it by means of the definitions
	\beq \label{hr-forms}
	\omega_{h_r}( \mathbb{X}_H , \, \cdot \, ) = \d f_{\hat H} \, ,
	\qquad
	g_{h_r} = ( \mathbb{Y}_H , \, \cdot \, ) = \d f_{\hat H} \, ,
	\eeq
where $f_{\hat H}$ is a smooth linear functions on $h_r$. 
These functions are determined by the expectation value of a Hamiltonian operator.
Let us consider Hamiltonian operators quadratic in the position and momentum of the general form
	\beq \label{Hamiltonian-operator}
	\hat{H} = \frac{1}{2} \left[ H_1 \, \hat{p}^2 + V(\hat{q} \, \hat{p} + \hat{p} \, \hat{q}) + H_2 \, \hat{q}^2 \right] \, ,
	\eeq
where $H_1$, $H_2$ and $V$ are real constants.
Thus, the expectation value of the Hamiltonian operator $\hat{H}$, defined as $f_{\hat{H}} :=  \Tr \{ \hat{\rho} \, \hat{H} \} $, is given by 
	\beq
	f_{ \hat H} =
	 \frac{1}{2} \left[ H_1 \, \langle \hat{p}^2 \rangle+ V \langle \hat{q} \, \hat{p} + \hat{p} \, \hat{q}\rangle 
	+ H_2 \langle \hat{q}^2 \rangle \right]\, 
	\eeq
and from the definitions \eqref{uncertainties} and \eqref{correlation} it is not difficult to find that	
	\beq \label{expectation-value-H}
	f_{\hat H} =  \frac{1}{2} \left[
	 H_1 \, \sigma^2_p + 2 \, V \sigma_{qp} + H_2 \sigma_q^2 
	 \right] \, .
	\eeq

Determining the Hamiltonian and gradient dynamics boils down to obtain the vector fields $\mathbb{X}_{H}$ and $\mathbb{Y}_{H}$ as defined in \eqref{hr-forms}. We can employ the coordinates adapted to the complex upper-half-plane defined in \eqref{C-coordinates} to express the symplectic form and the Riemannian metric
	\beq \label{C-forms}
	\omega_{h_r} 
	= \frac{ \i \,\hbar}{2} \frac{r^2}{(\bar{\mathcal{C}} - \mathcal{C})^2} 
	\d \, \bar{\mathcal{C}} \wedge \d \, \mathcal{C}
	\qquad
	\text{and}
	\qquad
	g_{h_r}	 
	= \frac{\hbar}{2} \frac{r^2}{(\bar{\mathcal{C}} - \mathcal{C})^2} 
	\d \, \bar{\mathcal{C}} \otimes_{\tiny \mbox{s}} \d \, \mathcal{C} \, ,
	\eeq
with complex structure
	\beq
	J_{h_r} = \frac{1}{\i}
	\left(
	\d \Ri \otimes \frac{\partial}{\partial \Ri } - \d  \bar{\Ri} \otimes \frac{\partial}{\partial  \bar{\Ri} }
	\right) \, .
	\eeq
The expectation value of the Hamiltonian operator $f_{\hat{H}}$ in \eqref{expectation-value-H} has the following expression in the $\mathcal{C}$ coordinate system	
	\beq
	f_{ \hat H}
	= \frac{ \i \,\hbar}{2} \frac{r}{\Ri - \bar{\Ri} }
	\left[
	H_1 \, \Ri \,  \bar{\Ri} + V \, (\Ri + \bar{\Ri}) + H_2
	\right]
	\eeq
where there is a clear dependence on the parameter $r$. 
We may now proceed to compute the Hamiltonian and the gradient vector fields considering that they take the following generic form	
	\beq \label{Pure-Vector-Fields}
	\mathbb{X}_{H} = \mathbb{X}_\Ri \frac{\partial}{\partial \Ri}+\mathbb{X}_{\bar{\Ri}} \frac{\partial}{\partial \bar{\Ri}}
	\quad
	\text{and}
	\quad
	\mathbb{Y}_{H} = \mathbb{Y}_\Ri \frac{\partial}{\partial \Ri}+\mathbb{Y}_{\bar{\Ri}} \frac{\partial}{\partial \bar{\Ri}} \ ,
	\eeq
respectively. 
Then, employing the definition in Eq.~\eqref{hr-forms} one may find that the components of these vector fields are given by
	\beq
	\mathbb{X}_\Ri = - \frac{2}{ \i \,\hbar} \frac{(\Ri - \bar{\Ri})^2}{r^2} \,
	 \frac{\partial f_H}{\partial \bar{\Ri}}
	= -\frac{1}{r} \left[ H_1 \, \Ri^2 + 2\, V \, \Ri + H_2 \right] \ ,
	\eeq
and
	\beq \label{Hyp-Grandient-like-Vector}
	\mathbb{Y}_\Ri = \frac{2}{\hbar} \frac{(\Ri - \bar{\Ri})^2}{ r^2 } \, 
	\frac{\partial f_H}{\partial \bar{\Ri}}
	= \frac{\i}{r} \left[ H_1 \, \Ri^2 + 2\, V \, \Ri + H_2 \right] \, .
	\eeq
Therefore, the symplectic evolution in these coordinates is described by the nonlinear Riccati equation
	\beq \label{Gaussian-Ric-Eq}
	\dot{\Ri} = - \frac{1}{r} \left[ H_1 \, \Ri^2 + 2\, V \, \Ri + H_2 \right] \, .
	\eeq   
	
On the other hand, the equations of motion in terms of Euclidean coordinates $(y^1,y^2, y^3)$ are given by the linear system of equation
	\begin{align}
	\dot{y}^1 & = \frac{1}{r} [(H_1 -H_2) \, y^2 - 2 \, V y^3] \, ,\\
	\dot{y}^2 & = \frac{1}{r} [(H_1 -H_2) \, y^3 + (H_1 + H_2) \, y^1] \, ,  \\
	\dot{y}^3 & = - \frac{1}{r} [(H_1 + H_2) \, y^1 + 2 \, V y^2] \, ,
	\end{align}
where the solutions of this linear system of equations correspond to the integral curves of the linear system of equations obtained from the Hamiltonian vector field
	\beq 
	\mathbb{X}_{H} = 
	\frac{1}{r} [(H_1 -H_2) \, y^2 - 2 \, V y^3] \frac{\partial}{\partial y^1} 
	+ \frac{1}{r} [(H_1 -H_2) \, y^3 + (H_1 + H_2) \, y^1] \frac{\partial}{\partial y^2} 
	- \frac{1}{r} [(H_1 + H_2) \, y^1 + 2 \, V y^2]  \frac{\partial}{\partial y^3} \, .
	\eeq


Let us now introduce the Poisson and Jordan brackets in $h_r$ and establish their relation to the algebraic structures of operators in $\D$. To this end, we start defining from $\omega_{h_r}$ the Poisson bracket for observables $f_{\hat{a}}$ and $f_{\hat{b}}$ associated to the operators $\hat{a}$ and $\hat{b}$, respectively. Thus, the Poisson bracket is defined by means of the symplectic form \eqref{C-forms} defined on each leaf $h_r$ by
	\beq 
	\{ f_{\hat a} , f_{\hat b} \}_{h_r} 
	= \omega_{h_r}(\mathbb{X}_b , \mathbb{X}_a ) \, .
	\eeq
Then, it can be shown by direct calculation that the Poisson bracket satisfies the following relation
	\beq \label{Lie-sl2}
	\{ f_{\hat a} , f_{\hat b} \}_{h_r}  = \frac{1}{r} \, f_{[[ \hat{a}, \hat{b} ]]} \, ,
	\eeq
where $[[\cdot,\cdot]]$ is the Lie-product defined in Eq. \eqref{Lie-Jordan-algebra}, where to obtain the last identity we have employed the basis $\{ \hat{L}^k \}$ in Eq. \eqref{SL-algebra-basis} satisfying the commutation relations \eqref{sl-Lie-product}.
On the other hand, the Jordan bracket is defined through the relation
	\beq 
	\langle \langle f_{\hat a} , f_{\hat b} \rangle \rangle_{h_r} 
	= g_{h_r}(\mathbb{Y}_a , \mathbb{Y}_b ) \, .
	\eeq

Hence, taking into account this definition, it can be shown by direct substitution that the Jordan bracket satisfies the following relation 	
	\beq  \label{quantum-Jordan-sl2}
	\langle \langle f_{\hat a} , f_{\hat b} \rangle \rangle_{h_r} = 
	\frac{1}{2} \, f_{\hat{a} \odot \hat{b}} - \frac{4}{\hbar \, r^2} \, f_{\hat a} \,  f_{\hat b} \, ,
	\eeq
where $\odot$ is the Jordan-product defined in Eq. \eqref{Lie-Jordan-algebra}, which for the basis $\{ \hat{L}^\mu \}$ obeys the relations \eqref{Jordan-Product-sl}.

To conclude this subsection, let us introduce a couple of bivector fields which will allow us to extend the Hamiltonian and gradient dynamics to the whole space $\mathbf{H}$. 
These two tensor fields will be relevant in introducing the GKLS dynamics in the following subsections. 
On the one hand, the skew-symmetric bivector field will permit to describe the Hamiltonian dynamics on $\mathbf{H}$ where a symplectic form cannot be defined; in this sense, this is a more general geometric field which can be used to define a Poisson bracket. 
On the other hand, the symmetric bivector field will serve for generalizing the gradient dynamics, although some redefinitions will be necessary to establish properly the GKLS dynamics.

The skew-symmetric bivector field in the $\Ri$ coordinates is given by the expression  
	\beq \label{pure-Poisson-Bivectors}
	\Lambda_{h_r} = \frac{2 \, \i}{\hbar } \frac{(\Ri - \bar{\Ri})^2}{r^2}
	\frac{\partial}{\partial \bar{\Ri}} \wedge \frac{\partial}{\partial \Ri}
	\eeq
and is such that the Poisson brackets can be defined as 
	\beq \label{2-bivectors}
	\{ f_{\hat a}, f_{\hat b} \}_{h_r} : = \Lambda_{h_r} (\d f_{\hat a}, \d f_{\hat b}) \, ,
	\eeq
while the symmetric bivector field takes the form
	\beq \label{Symmetrical-Bivectors} 
	G_{h_r} = \frac{2}{\hbar } \frac{(\Ri - \bar{\Ri})^2}{r^2}
	\frac{\partial}{\partial \bar{\Ri}} \otimes_{\tiny \mbox{s}} \frac{\partial}{\partial \Ri} \, ,
	\eeq
and the Jordan bracket is defined by the following relation
	\beq
	\langle \langle f_{\hat a}, f_{\hat b} \rangle \rangle_{h_r} : = G_{h_r} ( \d f_{\hat a} , \d f_{\hat b} ) \, .
	\eeq

Therefore, from the state point of view, we have established the symplectic evolution and the gradient vector field which are tangent to the manifolds $h_r$; however, to introduce the GKLS evolution, it is necessary to generalize these definitions to all the manifold $\mathbf{H}$. 
To achieve this, we must consider the \emph{observable point of view}, as we will see in the next section.


\subsection{Dynamical study of Gaussian state systems from an observable point of view} \label{subsec.3.2}

In the previous subsections, as well as in the q-bit case, we have seen that the symplectic and gradient dynamics generated by the corresponding vector fields preserve the degree of purity on each leaf $h_r$, that is, the parameter $r$ remains unchanged during this type of evolution. This means non-pure states with initial conditions described by a density matrix fulfilling \eqref{constrain-GS} preserve such a constraint under this dynamics. However, we aim to find a dynamical evolution for the states that does not preserve the degree of purity of the Gaussian state; in a geometric language we are looking for dynamics described by a vector field transverse to the leaves of constant $r$, that is, to the hiperboloids $\{h_r \}$. This evolution will take place in the manifold $\mathbf{H}$ and the change in the parameter $r$ will reflect the fact that the purity of states is changing in time.

Then, similarly to what has been done in the q-bit case, we shall establish the dynamics generated by the GKLS equation in geometrical terms, finding an affine vector field $\Gamma$ which accepts a decomposition in terms of a Hamiltonian dynamics, a gradient-like vector field and a Choi--Kraus vector field, as is expressed in Eq. \eqref{Affine-vector-field}.  Thus, to do so, we will extend the definition of Hamiltonian vector field and introduce the gradient-like vector field considering a description in terms of the \emph{observables} employing the Lie--Jordan structure $(\D, \odot, [[\cdot,\cdot]])$ of the space of observable operators.

From the \emph{observables} point of view, there is a one-to-one correspondence between each operator $\hat{a} \in \D$ and a function $\tilde{f}_{\hat{a}} := \Tr \{ \hat{a} \, \hat{\xi} \}$ on the dual space $\D^\ast$; therefore, the space of functions in $\D^\ast$ provides a realization of the Lie--Jordan algebra given in Eq. \eqref{Lie-Jordan-realizations}.

We may introduce a skew-symmetric and a symmetric $(2,0)$ tensor fields on $\D^\ast$ by means of its Lie and Jordan algebraic structures, respectively. At each point of $\D^\ast$ we have a cotangent space whose elements are the 1-forms $\d \tilde{f}_{\hat a}$ from the linear functions on $\D^\ast$ associated to any operator $\hat{a} \in \D$; then, we can define these bi-vector fields as
	\beq \label{Bivector-GDS}
	\tilde{\Lambda}(\d \tilde{f}_{\hat a}, \d \tilde{f}_{\hat b}) :=   \tilde{f}_{ [[\hat{a} , \hat{b}]] }
	\qquad
	\text{and} 
	\qquad
	\tilde{G}(\d \tilde{f}_{\hat a} , \d \tilde{f}_{\hat b}) := \, \tilde{f}_{\hat{a} \odot \hat{b}} \, ,
	\eeq
where $\tilde{f}_{\hat a}$ and $\tilde{f}_{\hat b}$ are functions on $\D^\ast$.
These tensor fields allow us to define a Hamiltonian vector field $\tilde{X}_{H}$ and a gradient-like vector field $\tilde{Y}_{b}$ on $\D^\ast$ as follows
	\beq \label{Definition-Vector-Fields}
	\tilde{X}_H = \tilde{\Lambda}( \d \tilde{f}_{\hat H}, \, \cdot \, )
	\qquad
	\text{and}
	\qquad
	\tilde{Y}_b = \tilde{G}( \d \tilde{f}_{\hat b}, \, \cdot \,) \, ,
	\eeq
where $\tilde{X}_{H}$ and $\tilde{Y}_{b}$ are associated to the operators $\hat{H}$ and $\hat{b}$, respectively.
Therefore, we obtain a Hamiltonian vector field associated to the Lie structure and a gradient like-vector field associated to the Jordan algebra structure. 

The Lie bracket of the set of generators defines the Lie algebra through the specification of its structure constants, then, for this basis the Lie product is such that
	\beq
	[[ \hat{L}^\mu , \hat{L}^\nu ]] = c^{\mu\nu}_\sigma \, \hat{L}^\sigma 
	\quad 
	\text{where} 
	\quad
	c_\sigma^{\mu\nu} = \begin{cases}
	0  & \text{for} \quad c_\sigma^{0 \nu}, \, \, c_\sigma^{\mu 0}, \, \, c_0^{\mu\nu} \, \\
	\\
	c^{k j}_{l} & \text{given by Eq.} \,\, \eqref{sl-Lie-product}  \, ,
	\end{cases}
	\eeq
where $k$, $j$ and $l$ denotes the values $1,2,3$ of $\mu$, $\nu$ and $\sigma$, respectively. 
Likewise, the Jordan product must fulfill the relations
	\beq \label{d-constants} 
	\hat{L}^\mu \odot \hat{L}^\nu = d_\sigma^{\mu\nu} \hat{L}^\sigma
	\quad 
	\text{with} 
	\quad
	d_\sigma^{\mu\nu} = \begin{cases}
	0  & \text{for} \quad d_l^{0 0} \quad \text{and} \quad d_l^{k j}  \, , \\
	\\
	g^{\mu \nu} & \text{for} \quad d_0^{\mu\nu} \, , \\
	\\
	\delta^{\mu}_\sigma & \text{for} \quad d_\sigma^{\mu 0} = d_\sigma^{0 \mu} \, .
	\end{cases}
	\eeq 
	
We want to define vector fields on $\D^\ast$ to describe the dynamical evolution of systems, then, it is convenient to introduce Cartesian coordinates on $\D^\ast$ associated to the basis $\{ \hat{L}^\mu \}$ of $\D$ by	
	\beq \label{cartesian-coordinates}
	f_{\hat{L}^\mu} = \Tr\{ \hat{L}^\mu \hat{\rho} \}
	= \frac{\hbar}{4} y^\mu \, .
	\eeq
where $\{y^k\}_{k=1,2,3}$ are directly connected to the second moments by the identities in \eqref{hr-to-Sr}.
Hence, given the cartesian coordinate system $\{ \frac{\hbar}{4} y^\mu \}$ we proceed to compute the coordinate expression of the tensor fields $\tilde{\Lambda}$ and $\tilde{G}$ by means of the definition \eqref{Bivector-GDS} obtaining 
	\beq \label{LyGtilde}
	\tilde{\Lambda} = \frac{4}{\hbar} \, c_\sigma^{\mu\nu} y^\sigma
	\frac{\partial}{\partial y^\mu} \wedge \frac{\partial}{\partial y^\nu}
	\qquad
	\text{and} 
	\qquad
	\tilde{G} = \frac{4}{\hbar} \, d_\sigma^{\mu\nu} y^\sigma 
	\frac{\partial}{\partial y^\mu} \otimes \frac{\partial}{\partial y^\nu} \, .
	\eeq
Given the Hamiltonian operator $\hat{H} = H_\mu \hat{L}^\mu$ and an arbitrary operator $\hat{b} = b_\mu \hat{L}^\mu$ their duals (expectation values) correspond to $\tilde{f}_{\hat{H}} = \frac{\hbar}{4} \, H_\mu y^\mu$ and $\tilde{f}_{\hat{b}} =\frac{\hbar}{4} \, b_\mu y^\mu$; then, the Hamiltonian and the gradient-like vector fields are 
	\beq \label{G-Vector-fields}
	 \tilde{X}_{H} = c_\sigma^{\mu\nu} H_\mu y^\sigma \frac{\partial}{\partial y^\nu}
	 \qquad
	\text{and} 
	\qquad
	\tilde{Y}_b = d_\sigma^{\mu\nu} b_\mu y^\sigma \frac{\partial}{\partial y^\nu} \, ,	
	\eeq
respectively.

The orbits obtained from the Hamiltonian and gradient-like vector fields must lie entirely on the space of quantum states $\mathbf{H}$ defined by the constraint $\rho(\hat{\mathbb{I}}) = \Tr\{ \hat{\rho} \, \mathbb{I} \} = 1$, that is, $\rho$ fulfills the definition of physical state. The space of quantum states is a convex subset of a closed affine subspace $\T$ of $\D^\ast$, then, in the case of systems described by a Gaussian density matrix, the dynamical trajectories that follow the quantum states are constrained to satisfy the Robertson--Schr\"odinger uncertainty relation
	\beq \label{RS-Inq}
	\sigma_{q}^2 \, \sigma_{p}^2 - \sigma^2_{qp} \geq \frac{\hbar^2}{4} \, .
	\eeq
Therefore, the affine subspace of quantum states is the subset of $\D^\ast$ defined as
	\beq 
	\T := \{ \tilde{f}_{\hat{a}} = a_\mu f_{\hat{L}^\mu} \in \mathcal{F}(\D^\ast) \, 
	| \,  g^{k j} \tilde{f}_{{\hat L}^k} \tilde{f}_{{\hat L}^j} \geq \frac{1}{4}\, (\tilde{f}_{{\hat L}^0})^2 \} \, ,
	\eeq 
which in terms of the coordinate system $\{\frac{\hbar}{4}y^\mu\}$ can be characterized by those points in $\D^\ast$ such that $y^0 = 2$. Then, it is possible to introduce the canonical immersion $i := \T \to \D^\ast$, such that $f_{\hat a} = \frac{\hbar}{2} a_0 + \frac{\hbar}{4}a_k y^k$ is the pullback of the linear function $\tilde{f}_{\hat a} = \frac{\hbar}{4} a_\mu y^\mu$, i.e., $f_{\hat a} = i^* \tilde{f}_{\hat a}$. 

Consequently, we can define symmetric and skew-symmetric tensor fields $\Lambda$ and $G$ on $\T$ by performing a reduction of the algebraic structures given in \eqref{Bivector-GDS}. 
Following a similar procedure as in the q-bit case, the space of functions $\mathcal{F}(\T)$ can be described as the quotient space $\mathcal{F}(\D^\ast)/\mathcal{I}$ where $\mathcal{I} \subset \mathcal{F}(\D^\ast)$ is the closed linear subspace of functions vanishing on $\T$. 
Now, in order to have an algebraic structure on $\mathcal{F}(\D^\ast)/\mathcal{I}$ the subspace $\mathcal{I}$ must be an ideal of $\mathcal{F}(\D^\ast)$. 
One may prove directly that for an element $\tilde{f}_{\hat a}$ in $\D^\ast$ and an element $\tilde{f}_{\hat b} = \left(1 - \frac{y^0}{2} \right) b_k y^k$ in $\mathcal{I}$ we have that
	\beq
	\tilde{\Lambda}(\d \tilde{f}_{\hat a}, \d \tilde{f}_{\hat b} ) = \frac{\hbar}{4} \left(1 - \frac{y^0}{2} \right) 
	 \, c^{k l}_m \, y^m (a_k b_l -b_k a_l) \, ,
	\eeq
which vanishes in $\T$, meaning that the Poisson product of $\tilde{f}_{\hat a}$ and $\tilde{f}_{\hat b}$ defined through $\tilde{\Lambda}$  is an ideal of $\mathcal{F}(\D^\ast)$.
Thus, because the differentials of the coordinate functions $\{\frac{\hbar}{4}y^k\}$ form a coordinate basis for the cotangent space at each point of $\T$, then, by means of the immersion $i$ we define the Poisson bivector field on $\T$ as
	\beq
	\Lambda(\d f_{\hat{L}^k}, \d f_{\hat{L}^j}) := f_{ [[ \hat{L}^k , \hat{L}^j]] } = c^{k j}_l f_{\hat{L}^l} \, ,
	\eeq  
or equivalently
	\beq
	\Lambda(\d y^k, \d y^j) = \frac{4}{\hbar}\, c^{k j}_l y^l\, ,
	\eeq 
hence, $\Lambda$ has the explicit form
	\beq  \label{y-Scr-Bivector}
	\Lambda := \frac{4}{\hbar} \, c^{k j}_l y^l
	\frac{\partial}{\partial y^k} \wedge \frac{\partial}{\partial y^j}
	=- \frac{4}{\hbar} \left( y^1 \, \frac{\partial}{\partial y^2} \wedge \frac{\partial}{\partial y^3}
	+ y^2 \, \frac{\partial}{\partial y^1} \wedge \frac{\partial}{\partial y^3}
	 - y^3 \, \frac{\partial}{\partial y^1} \wedge \frac{\partial}{\partial y^2}
	\right) \; .
	\eeq	
Given the reduced bivector field $\Lambda$ on $\T$ we define the Hamiltonian vector field $X_{\hat H} := \Lambda(\d f_{\hat H},\, \cdot \,)$, whose coordinate expression can be obtained considering the linear function $i^* \tilde{f}_{\hat H} = f_{\hat H} = \frac{\hbar}{2} H_0 + \frac{\hbar}{4} \, H_k  \, y^k $ associated to the Hamiltonian operator, this is, 
	\beq \label{Gauss-Ham-vec-field}
	X_H = c^{k j}_l H_k \, y^l \frac{\partial}{\partial y^j} =
	- (y^3 H_2 - y^2 H_3) \frac{\partial}{\partial y^1} 
	+ (y^1 H_3 + y^3 H_1) \frac{\partial}{\partial y^2} 
	- (y^1 H_2 + y^2 H_1) \frac{\partial}{\partial y^3} \, .
	\eeq	
	
An additional advantage of deducing the dynamics from an observable (algebraic) point of view is that a constant of motion can be obtained from the Casimir operator, which in this case is $\hat{C} = (\hat{L}^1)^2 - (\hat{L}^2)^2 - (\hat{L}^3)^2$, and whose expectation value gives the constant of motion 
	\beq 
	f_{\hat C} = (f_{L^1})^2 - (f_{L^2})^2 - (f_{L^3})^2 = \frac{\hbar^2}{4^2} [ (y^1)^2 - (y^2)^2 - (y^3)^2 ] \, .
	\eeq
This constant of motion allows to easily verify that $\Lambda(\d \, f_H , \d \, f_C) = 0$, therefore, the Hamiltonian vector field $X_{\hat H}$ is tangent to the manifolds $\{h_r \}$. Moreover, from this result it can be seen that the affine space is actually the manifold $\mathbf{H}$ defined in Eq. \eqref{Sold-Hyperboloid}.

In what refers to the gradient vector field $\tilde{Y}_{\hat b} = \tilde{G}(\d \tilde{f}_{\hat b}, \cdot\, )$, it is not difficult to show that $\tilde{Y}_{\hat b}(f_{\hat C}) \neq 0$; then, it can be shown that upon reduction of $\tilde{Y}_{\hat b}$ we will obtain a vector field $Y_{\hat b}$ that generates a dynamical evolution that is not necessarily restricted to the manifold $\mathbf{H}$; in particular, we are interested in vector fields that are tangent to the space of pure states. 
To amend this situation, we proceed to perform a reduction procedure for the symmetrical bivector $\tilde{G}$; however, 
it is not difficult to show that $\mathcal{I}$ \emph{is not an ideal} of $\mathcal{F}(\D^\ast)$ under this product because of
	\beq
	\tilde{G}(\d \tilde{f}_{\hat a}, \d \tilde{f}_{\hat b} ) 
	= \frac{\hbar}{4} \left( 1-\frac{y^0}{2} \right) (y^0 g^{kl} \, a_k b_l + a_0 \, b_k x^k)
	- \frac{\hbar}{8} \, a_0 y^0 \, b_k y^k 
	- \frac{\hbar}{8} \, a_k y^k \, b_n y^n \, .
	\eeq 
for $\tilde{f}_{\hat b} \in \mathcal{I}$ and an arbitrary $\tilde{f}_{\hat a} \in \mathcal{F}(\D^\ast)$.
Then, it is necessary to modify this symmetrical bivector. 
Following the procedure for the q-bit case and taking into account the result in \eqref{quantum-Jordan-sl2} one may start modifying the bivector by $\tilde{G} - \frac{4}{\hbar} \tilde{\Delta} \otimes \tilde{\Delta}$ with  $\tilde{\Delta} = y^\mu \frac{\partial}{\partial y^\mu}$; however,  it takes a straightforward calculation to show that on the affine space
	\beq
	\tilde{G}(\d \tilde{f}_{\hat a}, \d \tilde{f}_{\hat b} )
	- \frac{4}{\hbar} \tilde{\Delta}(\d \tilde{f}_{\hat a}) \tilde{\Delta}(\d \tilde{f}_{\hat b} )  \neq 0,
	\eeq
hence, its associated algebraic product is not in the subset $\mathcal{I}$.
Furthermore, it can be shown that there is not a linear combination of the tensor fields $\tilde{G}$ and $\tilde{\Delta} \otimes \tilde{\Delta}$ which defines a product such that $\mathcal{I}$ is an ideal. 
Then, it is necessary to find a fine-tuned definition for a tensor field that allows to properly accomplish the reduction process. A possibility is to slightly modify the $\tilde{G}$ and introduce a different vector field $\tilde{\Delta}$ to define the following tensor field
	\beq
	\tilde{\mathcal{R}} := \frac{1}{2}\, \mathcal{G} - \frac{4}{\hbar} \, \tilde{\Delta} \otimes \tilde{\Delta} \, ,
	\eeq
where now $\tilde{\Delta} = \frac{y^0}{4} \frac{\partial}{\partial y^0} +  y^k \frac{\partial}{\partial y^k}$ and 
	\beq 
	\mathcal{G} = \frac{4}{\hbar}\, \tilde{d}^{\mu\nu}_\sigma\, y^\sigma\, \frac{\partial}{\partial y^\mu} \otimes 				\frac{\partial}{\partial y^\nu} \, ,
	\eeq
with $\tilde{d}^{00}_0 = d^{00}_0 - \frac{3}{2 y^0}$ and all the other constants in \eqref{d-constants} remaining the same, i.e., $\tilde{d}^{\mu k}_0 = d^{\mu k}_0$ and $\tilde{d}^{\mu\nu}_k = d^{\mu\nu}_k$. Notice that the vector field $\tilde{\Delta}$ is a modification of the Euler-Liouville vector field. 
This tensor field allows us to define a product for the space of functions $\mathcal{F}(\D^\ast)$ for which $\mathcal{I}$ is an ideal with respect to it, in fact it is direct to check that
	\beq
	\tilde{\mathcal{R}}(\d \tilde{f}_{\hat a}, \d \tilde{f}_{\hat b} ) =  
	\left( 1-\frac{y^0}{2} \right) \left(\frac{\hbar}{8} y^0 g^{kl} \, a_k b_l + \frac{\hbar}{8} a_0 \, b_k x^k (1- y^0) 
	- \frac{5 \hbar}{16} \, a_k y^k \, b_n x^n
	\right)
	+\frac{3 \hbar}{32} \, b_k y^k \, \left( 1 - \frac{(y^0)^2}{4}\right) \, ,
	\eeq
which is an element of the ideal. 	
In particular, it will prove useful to have the tensor field $\tilde{\mathcal{R}}$ expressed in terms of the bivector field $\tilde{G}$ introduced in \eqref{LyGtilde} by
	\beq  \label{RtildeFText}
	\tilde{\mathcal{R}} = \frac{1}{2} \tilde{G} - \frac{3}{\hbar}\, \frac{\partial}{\partial y^0} \otimes \frac{\partial}{\partial y^0} - \frac{4}{\hbar} \tilde{\Delta} \otimes \tilde{\Delta}\, .
	\eeq 
In this manner, the gradient-like vector field associated to $\tilde{f}_{\hat{b}} = \frac{\hbar}{4} b_\mu y^\mu$ is
	\beq  \label{RtildeFTdfbext}
	\tilde{\mathcal{R}}(\d {\tilde f}_{\hat{b}} \, , \, \cdot \,) = \frac{1}{2}\tilde{Y}_{\hat b} 
	- \frac{3}{4} b_0 \frac{\partial}{\partial y^0} 
	- \left(b_0 \frac{y^0}{4} + b_k y^k\right) \tilde{\Delta} \, .
	\eeq
where the gradient-like vector field $\tilde{Y}_{\hat b}$ is defined in \eqref{Definition-Vector-Fields}.
	
Once we have defined the symmetrical product $\tilde{\mathcal R}$, we may proceed to perform the reduction by means of the immersion $i^* \tilde{f}_{\hat H} = f_{\hat H} = \frac{\hbar}{2} H_0 + \frac{\hbar}{4} \, H_k  \, y^k$, to obtain the  reduced symmetric bivector $\mathcal{R}$ defined by
	\beq
	\mathcal{R}(\d f_{\hat{L}^k}. \d f_{\hat{L}^j}) =
	\frac{1}{2} f_{\hat{L}^k \odot \hat{L}^j} - \frac{4}{\hbar} \, f_{\hat{L}^k} \,  f_{\hat{L}^j} \, ,
	\eeq 
such that in Cartesian coordinates takes the form
	\beq
	\mathcal{R}
	= \frac{4}{\hbar}  \left( g^{k j} - y^k y^j\right) \frac{\partial}{\partial y^k} \otimes \frac{\partial}{\partial y^j} \, ,
	\eeq 
where we have denoted $\Delta = y^k \frac{\partial}{\partial y^k}$.	
Now, by means of $\mathcal{R}$ we can define the gradient-like vector field associated to the linear function $f_{\hat{b}} = \frac{\hbar}{2} b_0 + \frac{\hbar}{4} b_k y^k$, which results in
	\beq \label{Yb}
	Y_b := \mathcal{R} (\d f_{\hat b}, \, \cdot \,) 
	= \left(g^{kj}  -   y^k y^j \right) b_k  \, \frac{\partial}{\partial y^j} \, ,
	\eeq  
where it is important to note that the last term in this gradient-like vector field is a quadratic term with respect to the Cartesian coordinate system. 

To finalize this subsection, we want to verify that the gradient-like vector field $Y_b$ generates a dynamic evolution that is transverse to the leaves $h_r$ but tangent to $H ^2$ when $r=1$, that is, when the initial conditions are those of a pure state; this means that the states reached from a dynamic evolution dictated by $Y_b$, regardless of the initial conditions,  are contained within $\mathbf{H}$. To do so, we compute the Lie derivative of $f_{\hat{C}}$ along the direction of the gradient-like vector field $Y_b$ obtaining
	\beq
	\pounds_{Y_{b}} \, f_{\hat{C}} = \frac{\hbar^2}{2}\left(1 - r^2 \right) b_k y^k \, .
	\eeq 
Thus, we observe that the Lie derivative is different from zero if and only if $r \neq 1$; therefore, the gradient-like vector field $Y_b$ is transversal to the leaves $h_r$ and is tangent to the hyperboloid $H^2$.
Moreover, when $r = 1$, the gradient-like vector field $Y_b$ is identical to the gradient vector field $\mathbb{Y}_b$, meaning that dynamics generated by $Y_b$ does not change the purity of the states, because a pure state remains pure under its evolution.


\subsection{GKLS dynamics for Gaussian states} \label{subsec.3.4}

In this subsection we construct a geometric description of the GKLS dynamics of physical systems described by Gaussian states. This is done introducing a vector field which is defined from the GKLS master equation; this vector field is transverse to the foliation $\{h_r\}$ and can describe the evolution of a pure state into a non-pure one.

Therefore, although $\hat{\xi}$ is not a density matrix, we are able to establish a one-to-one connection between operators and their expectation values which is equivalent, in the sense that it reproduces the same mapping between $\D$ and $\D^\ast$, as the density matrix would do. 
Then, once established a map between $\D$ and $\D^\ast$ using the matrix representation of the Jordan--Lie algebra, we may now introduce the GKLS generator $\mathbf{L}$ acting on $\hat{\xi}$
	\beq \label{L-generator2}
	\mathbf{L}(\hat{\xi}) =  - \frac{\i}{\hbar} [\hat{H}, \hat{\xi} ]_{-}
	- \frac{1}{2} \sum^3_j [ \hat{v}_j^\dagger \, \hat{v}_j , \hat{\xi} ]_{+}
	+\sum^3_A \hat{v}_j \, \hat{\xi} \, \hat{v}_j^\dagger \, ,
	\eeq
where $\hat{H} \in \D$ is the Hamiltonian operator and $\hat{v}_A$ an arbitrary element of $\D$. 
This generator can be expressed in terms of the Lie and the Jordan products
	\beq  \label{L-gen-lie-jordan}
	\mathbf{L}(\hat{\xi}) =  - [[ \hat{H}, \hat{\xi} ]]
	- \frac{\hbar}{2} \, \hat{V} \odot \hat{\xi} + \hat{\mathcal{K}}( \hat{\xi} ) \, ,
	\eeq
where 
	\beq
	\hat{V} =  \sum_j \hat{v}_j^\dagger \hat{v}_j
	\eeq 
and the completely positive map is
	\beq  \label{positive-map}
	\hat{\mathcal{K}}( \hat{\xi} ) = \sum_j \hat{v}_j \, \hat{\xi} \, \hat{v}_j^\dagger \, .
	\eeq
Let us now, in analogy with the analysis performed for the q-bit system, propose that the vector field on $\D^\ast$ associated with the $\mathbf{L}$ generator takes the form
	\beq \label{Gauss-Gamma-Tilde}
	\tilde{\Gamma} = \tilde{X}_{H} -  \frac{\hbar}{2} \, \tilde{Y}_{V} + \tilde{Z}_{\mathcal{K}} \, ,
	\eeq 
where $\tilde{X}_{H}$ is the Hamiltonian vector field $\tilde{\Lambda}( d \tilde{f}_{\hat{H}}, \cdot \ ) = \tilde{X}_{H} $, $\tilde{Y}_{V}$ is the gradient-like vector field $\tilde{G}( \d \tilde{f}_{\hat{V}}, \cdot \ ) = \tilde{Y}_{V}$ and $\tilde{Z}_{\mathcal{K}} $ is the vector field associated with the completely positive map $\hat{\mathcal{K}}$.

To find the explicit form of the GKLS vector field we consider the master equation for $\hat{\xi}$ in terms of the Lie and Jordan products and the completely positive map \eqref{L-gen-lie-jordan}; expressing the Hamiltonian operator as $\hat{H} = H_\mu \, \hat{L}^\mu$, the Lie product of $\hat{H}$ and $\hat{\xi}$ is
	\beq
	[[ \hat{H}, \hat{\xi} ]]
	=  c_\sigma^{\mu \nu} H_\mu \, y_\nu \, \hat{L}^\sigma \, ,
	\eeq
while the Jordan product of $\hat{V} = V_\mu \hat{L}^\mu$ and $\hat{\xi}$ is expressed as
	\beq 
	\hat{V} \odot \hat{\xi} =  d^{\mu\nu}_\sigma V_\mu y_\nu \hat{L}^\sigma \, .
	\eeq
Finally, we compute explicitly the completely positive map $\mathcal{K}( \hat{\xi} )$ in terms of the generators of $\sl(2, \R)$. Expressing it as
	\beq  \label{K-form}
	\hat{\mathcal{K}}( \hat{\xi} ) = \frac{4}{\hbar} \, \mathcal{K}_\mu \, \hat{L}^\mu
	=  \frac{4}{\hbar} \, \left\langle L_\mu \, , \, \hat{\mathcal{K}}( \hat{\xi} ) \right\rangle \, \hat{L}^\mu  \, ,
	\eeq
where the coefficients $\mathcal{K}_\mu$ are given by
	\beq \label{Kmu}
	\mathcal{K}_\mu
	= \left\langle L_\mu \, , \, \hat{\mathcal{K}}( \hat{\xi} ) \right\rangle 
	= \frac{1}{2\hbar} \Tr \left\{ \sum_A  \hat{v}_A \hat{L}^\nu \hat{v}_A^\dagger \hat{L}^\alpha \right\} y_\nu 				g_{\alpha\mu} \, , 
	\eeq
which can also be expressed as $\mathcal{K}_\mu = \mathcal{K}^\nu{}_\mu\, y_\nu$ 
where
	\beq \label{Kud}
	\mathcal{K}^\nu{}_\mu = \frac{1}{2\hbar} 
	\Tr \left\{ \sum_A \hat{v}_A \hat{L}^\nu \hat{v}_A^\dagger  \hat{L}^\alpha \right\} g_{\alpha\mu} \, .
	\eeq
Therefore, we finally have that
	\beq \label{op-K-final}
	\hat{\mathcal{K}}(\hat{\xi}) =  \frac{4}{\hbar} \mathcal{K}^\mu{}_\nu\, y_\mu \, \hat{L}^\nu \, .
	\eeq

Once that all the linear maps in \eqref{L-gen-lie-jordan} are expressed in terms of the $\sl(2, \R)$ generators, we are in the position to associate a linear vector field on $\D^\ast$ to each one of them. 
To do so, we recall that for a linear map $\hat{A}$ acting on $\hat{\xi}$ given by $\hat{A}(\hat{\xi}) = \mathcal{A}^\mu{}_\nu y_\mu \hat{L}^\nu$ we can associate a vector field on the dual $\D^\ast$ given by 
	\beq
	Z_{A} = \mathcal{A}_\mu{}^\nu y^\mu \frac{\partial}{\partial y^\nu} \, ,
	\eeq
where $\mathcal{A}_\mu{}^\nu \equiv g_{\alpha\mu}g^{\beta\nu} \mathcal{A}^\alpha{}_\beta$.
Then, the vector field associated to the GKLS generator $\mathbf{L}$ in \eqref{L-gen-lie-jordan} is given by
	\beq 
	\tilde{\Gamma} = - Z_{[[H, \xi ]]} - \frac{\hbar}{2} \, Z_{V \odot \xi } 
	+ \tilde{Z}_{\mathcal{K}} \, ,
	\eeq 
where each vector field that composes $\tilde{\Gamma}$ is given by
	\begin{equation}
	Z_{[[H , \xi]]} = - \, c_\sigma^{\mu \nu} \, H_\mu \, y^\sigma \frac{\partial}{\partial y^\nu} \, ,
	\quad
	Z_{V \odot \xi } = d_\sigma^{\mu \nu} \, V_\mu \, y^\sigma \frac{\partial}{\partial y^\nu} \quad \text{and} \quad
	\tilde{Z}_{\mathcal{K}} 
	= \frac{4}{\hbar} \, \mathcal{K}_\mu{}^\nu \, y^\mu \, \frac{\partial }{\partial y^\nu} \, ,
	\end{equation}
where to arrive to this result we have used that $L_\nu =  g_{\nu \eta} \hat{L}^\eta$ and $y_\nu \equiv g_{\nu \eta} y^\eta$, as well as $g_{\nu\alpha}g^{\sigma\beta} c^{\mu\nu}_\sigma = -c^{\mu\beta}_\alpha$, $g_{\nu\alpha}g^{\sigma\beta} d^{\mu\nu}_\sigma = d^{\mu\beta}_\alpha$ and $\mathcal{K}_\mu{}^\nu = g_{\mu\alpha}g^{\nu\beta} \mathcal{K}^\alpha{}_\beta$.
Then, comparing with the coordinate expression of the vector field in Eq. \eqref{G-Vector-fields} it is direct that $Z_{[[H, \xi ]]} = - \tilde{X}_H$ and $Z_{V \odot \xi }  = \tilde{Y}_V$.
Then, we have obtained an expression in terms of the coordinate basis for the vector field $\tilde{\Gamma}$ in Eq. \eqref{Gauss-Gamma-Tilde} associated to the GKLS master equation.	

Now, we would like to consider the restriction of $\tilde{\Gamma}$ to the the affine space $\T$ and obtain a vector field $\Gamma$, where such affine space generates a dynamic evolution whose orbits lie entirely on the space of physical quantum states. 
To accomplish this \textcolor{magenta}{restriction} by means of the immersion $i:= \T \to \D^\ast$, let us first express $\tilde{\Gamma}$ as
	\beq
	\tilde{\Gamma} = \tilde{X}_{H} - \hbar \, \left[ \frac{1}{2}\tilde{Y}_{\hat{V}} 
	- \frac{3}{4} V_0 \frac{\partial}{\partial y^0} 
	- (V_0 \frac{y^0}{4} + V_k y^k )\tilde{\Delta}\right] 
	+   \tilde{Z}_{\hat{\mathcal{K}}} 
	- \frac{3\hbar}{4} V_0 \frac{\partial}{\partial y^0} 
	- \hbar \left(V_0 \frac{y^0}{4} + V_k y^k\right) \tilde{\Delta} \, ,
 	\eeq 
at this point it is convenient to substitute the vector field $\tilde{Z}_{\hat{\mathcal{K}}(\hat{\xi})}$ expressed as follows
	\beq 
	\tilde{Z}_{\mathcal{K}} =  2 \, \tilde{f}_{\hat{V}} \, \frac{\partial }{\partial y_0}
	+ \frac{4}{\hbar} \, \mathcal{K}_\mu{}^k \, y^\mu \, \frac{\partial }{\partial y^k} \, ,
	\eeq
and then setting $y^0 = 2$ we obtain the vector field $\Gamma$ as
	\beq
	\Gamma = X_{H} + Y_{b} + Z_{\mathcal{K}} \, ,
	\eeq
by setting $\hat{b = - \hbar V}$ and where the Choi-Kraus vector field is identified as
	\beq  \label{Choi-Kraus-gaussian}
	Z_{\mathcal{K}}= \frac{4}{\hbar} \mathcal{K}_\mu{}^k y^\mu \frac{\partial}{\partial y^k}  
	-  \hbar \left(\frac{V_0}{2} +  V_k y^k \right) \Delta \, ,
	\eeq
with $\Delta = y^k \frac{\partial}{\partial y^k}$ is the Euler-Liouville vector field on $\T$ and $Y_{V}$ is the reduced gradient-like vector field given by Eq. \eqref{Yb}.
It is convenient to have an explicit formula for the coefficients $\mathcal{K}_\mu{}^\nu$, thus, from \eqref{Kud} we obtain
	\beq  \label{Kdu}
	\mathcal{K}_\mu{}^k =\frac{1}{2\hbar} 
	\Tr \left\{ \sum_A \hat{v}_A {\hat L}^\alpha \hat{v}_A^\dagger {\hat L}^k \right\} g_{\alpha \mu}  \, . 
	\eeq

We can express the GKLS vector field $\Gamma$ in Cartesian coordinates considering the expectation values  $f_{\hat{H}} = \frac{\hbar}{2}H_0 + \frac{\hbar}{4} H_k y^k$ for the  Hamiltonian operator and $f_{\hat{V}} = \frac{\hbar}{2}V_0 + \frac{\hbar}{4} V_k y^k$.
Then, finally, we obtain the result we were looking for, the GKLS vector field is given by the following expression
	\beq \label{Gauss-GKLS-Vec}
	\Gamma = c^{k j}_l H_k \, y^l \frac{\partial}{\partial y^j}
	-  \hbar\, g^{kj} \, V_k \, \frac{\partial}{\partial y^j}
	+ \frac{4}{\hbar} \, \mathcal{K}_\mu{}^j \, y^\mu \, \frac{\partial }{\partial y^j} 
	-  \frac{\hbar}{2}V_0 \Delta \, ,
	\eeq	
which is a linear vector field because the nonlinear terms in $Z_{\hat{\mathcal{K}}}$ and $Y_{\hat{V}}$ cancel out in the sum.


\subsection{Damping phenomena for the harmonic oscillator dynamics} \label{subsec.3.5}

In this subsection, we analyze, as an example, the GKLS dynamics considering the fiducial state described by the Gaussian density matrix \eqref{Gaussian-Density-Matrix}. Let us consider as the conservative system the harmonic oscillator with its Hamiltonian operator given as
	\beq \label{Harm-Osc}
	\hat{H}_{\rm{Osc}} = \frac{1}{2}(\hat{p}^2 + \hat{q}^2) = 2 \, \hat{L}^1 \, ,
 	\eeq
where we have set all the parameters to unity, i.e., $m=1$ and $\omega = 1$. Thus, from \eqref{Gauss-Ham-vec-field} the Hamiltonian vector field has the form
	\beq
	X_H = 2 \left(
	y^3  \, \frac{\partial }{\partial y^2} - y^2  \, \frac{\partial }{\partial y^3}
	\right) \, .
	\eeq

To introduce dissipation, let us consider first the operator $\hat{v}_1 = \frac{\sqrt{\gamma}}{\hbar} \, {\hat L}^1$ and hence we have that $\hat{V} = \frac{\gamma}{2 \, \hbar} \hat{L}^0$ and accordingly the gradient-like vector field definition, in Eq. \eqref{Yb}, $Y_{V}$ vanishes. Therefore, the Choi-Kraus vector field $Z_{\mathcal{K}}$ is the only term introducing dissipation into the system. To calculate $Z_{\mathcal{K}}$ we first need to compute the coefficients $\mathcal{K}_\mu{}^\nu$ for $\hat{v}_1$ which can be obtained from the expression
	\beq
	\mathcal{K}_\mu{}^\nu = \frac{\gamma}{2\hbar^3}\, g_{\mu\alpha} \, \Tr \left\{ \hat{L}^1 \hat{L}^\alpha \hat{L}^1 \hat{L}^\nu \right\} \, .
	\eeq
Then, from \eqref{Choi-Kraus-gaussian} we find the Choi-Kraus vector field
	\beq \label{Choi-Krauss-osc1}
	Z_{\mathcal{K}} = - \frac{\gamma}{2}\left( y^2 \frac{\partial}{\partial y^2} + y^3 \frac{\partial}{\partial y^3} \right) \, .
	\eeq
Finally, we obtain the GKLS vector field 
	\beq \label{GKLS-Gauss-Ex-1}
	\Gamma = X_{H} + Z_{\mathcal{K}} = -\left( \frac{\gamma}{2}y^2 - 2 y^3\right)\frac{\partial}{\partial y^2}  -\left( \frac{\gamma}{2}y^3 + 2 y^2\right)\frac{\partial}{\partial y^3} \, .
	\eeq	
Then, the system of equations of motion is
	\begin{align}
	\dot{y}^1 &= 0 , \nonumber \\
	\dot{y}^2 &= - \frac{\gamma}{2}y^2 + 2y^3 , \\
	\dot{y}^3 &= -\frac{\gamma}{2}y^3 - 2y^2 , \nonumber 
	\end{align}
and its solution for initial conditions at $t=0$, denoted as $(y^1_0, y^2_0, y^3_0)$, is described by harmonic functions with the damping factor modulated by $\gamma$ 
	\begin{align}
	y^1(t) &= y^1_0 , \nonumber \\
	y^2(t) &= e^{-\frac{\gamma}{2}t} \left(y^2_0 \cos 2t + y^3_0 \sin 2t\right) , \\
	y^3(t) &= e^{-\frac{\gamma}{2}t} \left(y^3_0 \cos 2t - y^2_0 \sin 2t\right) \, . \nonumber
	\end{align}
	
\begin{figure}[! h t]
	\centering
	\includegraphics[width = 7.5 cm]{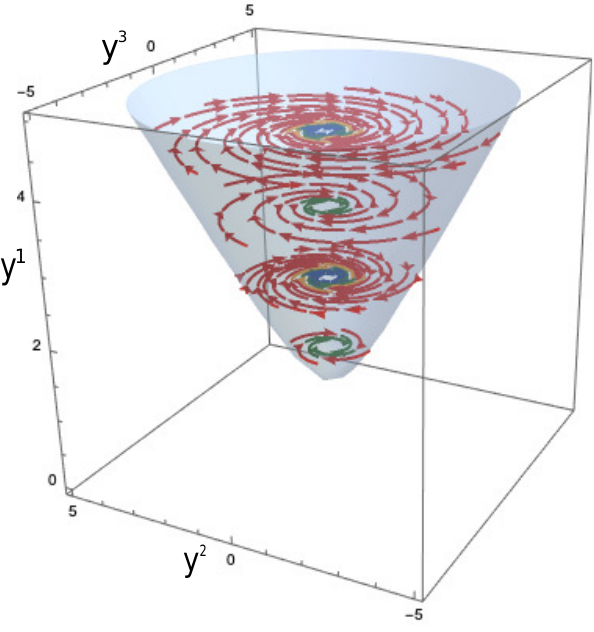}
	\caption{The GKLS vector field for a harmonic oscillator system with Hamiltonian $\hat{H}_{\tiny \mbox{Osc}} = 2 \, 		\hat{L}^1$ and a dissipative term introduced through the operator $v_1 = \frac{\sqrt{\gamma}}{\hbar} \, \hat{L}^1$ is 		displayed.
	Here we have the frequency of oscilation $\omega = 1$ and we consider the damping parameter 
	$\gamma = 1$. }
	\label{Fig-5}
	\end{figure}		
	
The GKLS vector field is plotted in Fig.~\ref{Fig-5} where the damping parameter has been set as unity, $\gamma = 1$. In this figure we observe that, regardless the initial conditions, the vector field converges asymptotically to the line $(y^1 , 0 , 0)$, which is a singular line for $\Gamma$. Moreover, we see that the pure coherent state given by the point $(1,0,0)$ will not be affected by this dynamics.

\vsp

As a second example of damping, we consider again the harmonic oscillator system in Eq. \eqref{Harm-Osc}, but now the dissipation term is given by 
	\beq
	\hat{v}_1 = \sqrt{\gamma} \hat{K}_{+} \, ,
	\eeq
where the operators $\hat{K}_+$ and $\hat{K}_-$ are given in terms of the $\sl(2,\R)$ generators as
	\beq
	\hat{K}_{+} = \frac{1}{\i \hbar} \left( \hat{L}^3 + \i \, \hat{L}^2 \right)
	\quad 
	\text{and}
	\quad
	\hat{K}_{-} = \frac{1}{\i \hbar} \left( \hat{L}^3 - \i \, \hat{L}^2 \right) \, ,
	\eeq	
and are such that $\hat{K}_{-}  = \hat{K}_{+}^\dagger$. 	
Thus, by direct calculation we find that $\hat{V} = \gamma \hat{K}_{-} \,\hat{K}_{+} = \frac{\gamma}{\hbar}(\hat{L}^0 - \hat{L}^1)$ and then, the gradient-like vector field can be computed by means of the result in \eqref{Yb}  
	\beq
	Y_{V} =  \frac{\gamma}{\hbar} \left[ ((y^1)^2 - 1)\frac{\partial}{\partial y^1} +  y^1 y^2 \frac{\partial}{\partial y^2} + y^1 y^3 \frac{\partial}{\partial y^3} \right] \, .
	\eeq 
From \eqref{Choi-Kraus-gaussian} and  \eqref{Kdu} we find that the Choi-Kraus term is 
	\beq \label{Choi-Kraus-2}
	Z_{\mathcal{K}} = \gamma  \left( (y^1)^2  - y^1 + 1 \right)\, \frac{\partial}{\partial y^1} +  \gamma \, \left( y^1 y^2 - 		\frac{y^2}{2} \right)\, \frac{\partial}{\partial y^2}  + \gamma \, \left( y^1 y^3 - \frac{y^3}{2} \right)\, \frac{\partial}{\partial 		y^3} \, .
	\eeq
Finally, the dynamical evolution is determined by the GKLS vector field $\Gamma = X_{H} - \hbar \, Y_{V} + Z_{\mathcal{K}}$, whose Cartesian coordinate expression is
	\beq \label{GKLS-osc2}
	\Gamma = -\gamma \left(y^1 - 2\right)\, \frac{\partial}{\partial y^1} 
	- \left( \frac{\gamma}{2} \, y^2 -  2 y^3   \right)\, \frac{\partial}{\partial y^2} 
	-  \left( \frac{\gamma}{2} \, y^3 + 2 y^2 \right)\, \frac{\partial}{\partial y^3} \, .
	\eeq

\begin{figure}[! t]
	\centering
	\includegraphics[width = 7.5 cm]{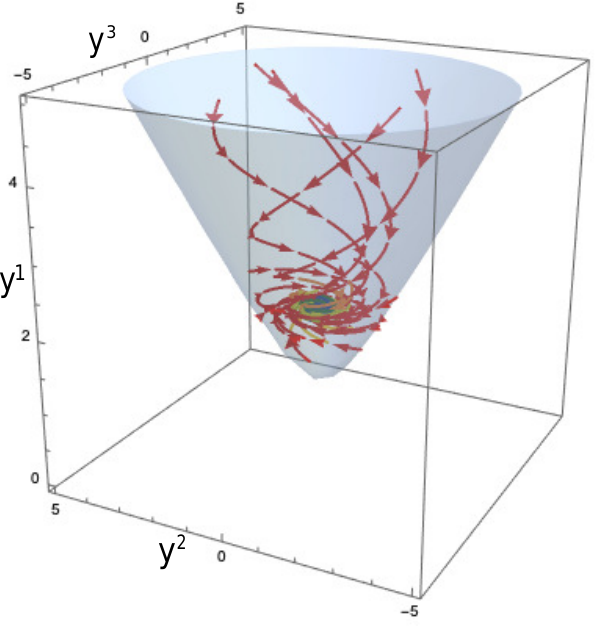}
	\caption{In this figure we plot the GKLS vector field for a Hamiltonian oscillator system 
	$\hat{H}_{\tiny \mbox{Osc}} = 2 \, \hat{L}^1$ with dissipative part introduced through
	the operator $\hat{v}_1 = \sqrt{\gamma} \, \hat{K}_{+}$. 
	Here we have considered the Cartesian coordinates 
	$(y^1, y^2, y^3) \in \mathbf{H}$ and the damping parameter $\gamma = 1$. }
	\label{Fig-6}
	\end{figure}

Notice that the components in the direction of $y^2$ and $y^3$ are equal to those in the GKLS vector field in \eqref{GKLS-Gauss-Ex-1} in the previous example; however, in this case, there is a non-vanishing component in the $y^1$ direction.	
The integral curves of this vector field are solutions to the linear system of equations	
	\begin{align}
	\dot{y}^1 & =   -\gamma \, (y^1 - 2) \, , \nonumber \\
	\dot{y}^2 & = -\frac{\gamma}{2}\, y^2 + 2 y^3 ,   \\
	\dot{y}^3 & = -\frac{\gamma}{2}\, y^3 - 2 y^2 \, ,  \nonumber
	\end{align} 
and are given by
	\begin{align}
	y^1(t) &= (y^1_0 - 2) e^{-\gamma t} + 2 , \nonumber \\
	y^2(t) &= e^{-\frac{\gamma}{2} t} \left(y^2_0 \cos 2t + y^3_0 \sin 2t \right) , \\
	y^3(t) &= e^{- \frac{\gamma}{2} t} \left(y^3_0 \cos 2t - y^2_0 \sin 2t \right) \, , \nonumber
	\end{align}
where again $(y^1_0,\, y^2_0,\, y^3_0)$ are initial conditions at $t=0$.
The GKLS vector field for this case is displayed in Fig. \ref{Fig-6}; here we observe that there is a unique singular point at $(2, 0 , 0)$. For any initial condition, every solution converges asymptotically to this singular point.

\section{Conclusions and perspectives} \label{sec.4}

In this work, we have obtained in detail the GKLS vector field for systems described by Gaussian states. 
In the first part, we have reviewed thoroughly the case of two-level systems, the q-bit, to introduce the concepts and procedures presented in \cite{Ciaglia-2017}, which are necessary to accomplish the same task for the dissipative dynamics of the Gaussian case. 
Although the two cases can be worked out similarly some differences must be taken into account.
In both cases, we start by introducing the space of quantum states and recognize these as parametrized spaces, that can be described as a manifold with boundary. 
The purity condition $\rho^2 = \rho$ determines the boundary of the manifold of states, while the non-pure condition $\Tr\{ \rho^2 \} < 1$ its interior, that is, the boundary is the space of pure states and the open interior describes non-pure ones. 
As a first distinct feature, the manifold representing the quantum states for the q-bit is compact, while in the case of the Gaussian system is open, where the isospectral submanifolds are hyperboloids \eqref{Foliation-H} instead of Bloch spheres.

To describe a dynamical evolution that allows a change in the degree of purity of the states, we have determined the GKLS vector field. 
The integral curves of this vector field are, in general, transversal to the foliation, making it possible to evolve from a pure or non-pure state to another non-pure one. 
This was done consistently, imposing that the orbits of this vector field were constrained to remain in the space of quantum states.  
 To find the GKLS vector field in the Gaussian case, we followed closely the construction for the $n$-levels systems presented in \cite{Ciaglia-2017}. 
 Thus, we considered the description of the space of quantum states from an observable point of view, which offers the advantage of endowing the space of observables with a Lie-Jordan algebra structure. 
 In particular, for the Gaussian density matrix case, we restricted our study to operators quadratic in the position and momentum operators.
 
The space of observables, that is, the space of real functions on the dual of the Lie-Jordan algebra provides a realization of the Lie-Jordan algebra of operators, thus, it is possible to define geometric structures on the space of observables from these algebraic products. 
Namely, associated with the Lie product it is possible to define a (skew-symmetric) Poisson bivector field and to the Jordan product a corresponding symmetrical bivector. 
The Poisson bivector field endows the space of states with a Poisson structure and a Hamiltonian vector field can be defined for the entire space, generalizing the symplectic structure, in particular, for the case of odd-dimensional spaces. 
On the other hand, the symmetric bivector field defines a gradient-like vector field that is transverse to the leaves of the foliation. 
To constraint the orbits associated with the Hamiltonian and gradient-like vector fields to the space of quantum states, it is necessary to define such tensors in this space employing a reduction procedure.  
This reduction allows us to find a new symmetric bivector which yields a vector field that is transverse to the foliation, but it is tangent to the leaf of pure states. 

The orbits of the vector field defined from this modified bivector defines a dynamics such that a pure state remains pure under this evolution. To amend this, a vector field associated with a completely positive map from the dual space to itself is introduced, defining in this way the GKLS vector field associated with the corresponding GKLS operator.  
Obviously, in the Gaussian case, the GKLS vector field is different from the one for the q-bit. 

Therefore, we have shown that, in the case of states described by a Gaussian state, the GKLS dynamics also admits a \emph{decomposition principle}, i.e. a conservative-Hamiltonian part as a reference dynamics, while the sum of the gradient-like and the Choi--Kraus vector fields are considered as ``perturbation terms'' associated with dissipation.
In this sense, we have seen that the very concept of dissipation is not associated with the GKLS vector field itself, but rather, with the decomposition of this vector field in terms of the relevant geometrical structures: the Hamiltonian vector field, the like-gradient vector field and the Choi--Kraus vector field.

There are still many investigations to pursue in the study of the evolution of GKLS for Gaussian density matrices. 
For instance, an immediate question is whether this procedure can be generalized to address systems with more dimensions. 
The generalization of this procedure to $n$-levels has been already obtained in Ref. \cite{Ciaglia-2017}, while the general form of the Gaussian density matrix with a statistical mixture for $n$-dimensions is well known and has been constructed employing a generalization of the so-called covariance matrix, see \cite{Ferraro-2005} and references therein. 
Thus, taking into account these results one could proceed directly to construct the GKLS vector field for more degrees of freedom. 
This will be presented in a forthcoming work.

Another important generalization of our procedure is to consider the Gaussian states with non-vanishing first moments. This implies considering observables at most quadratic in position and momentum operators, i.e., one may include linear operators such as the creation and annihilation operators. 
This will be studied in a future work.

Finally, in this work, we have only considered as a fiducial state the Gaussian density matrix, so a natural question is whether different fiducial states can be chosen. 
In particular, it might be possible to consider the following Wigner function. Let us define
	\beq 
	I(q,p,t) : = \frac{4}{\hbar^2 \, r^2} \left[
	\sigma_{p}^2 q^2 
	- 2 \sigma_{qp} q p
	+ \sigma_{q}^2 p^2 \right] \, ,
	\eeq
then, one may introduce the following family of Wigner functions
	\beq
	W_n(q , p) = \frac{(-1)^n}{\pi \, \hbar \, r} e^{-\frac{1}{2} I(q,p,t)} L_n[ I(q,p,t) ] \, ,
	\eeq
where $L_n$ denotes the Laguerre polynomials. 
Then, $r=1$ corresponds to Wigner functions associated with the Fock states; however,  for $r \neq 1$ these states satisfy the Robertson--Schr\"odinger uncertainty relation
	\beq 
	\sigma_{q}^2 \, \sigma_{p}^2  -  \sigma^2_{qp} = \frac{\hbar^2 \, r^2}{4} \, ,
	\eeq
and the purity condition 
	\beq 
	\Tr\{ \hat{\rho}^2 \} = \frac{1}{r} \, .
	\eeq
Notice that, $n = 0$, returns the Gaussian density matrix studied in this work. 
This set of states has the same space of parameters analyzed for the Gaussian density matrix and then our results might be applied directly, i.e., the GKLS vector field obtained in \eqref{Gauss-GKLS-Vec} could be used with fiducial states in the set $\{ W_n(q , p) \}$. 
These results will be studied in future contributions.

Let us indicate the main points of our construction.
We have selected  the subset of matrices of $\sl(2,\R)$ which may be considered as covariance matrices for Gaussian states associated with the Hilbert space on the real line, this is what we may call generalized coherent states.
Orbits of the coadjoint action of $SL(2,\R)$, analog of isospectral orbits for the q-bit case, orbits of $SU(2)$, are here characterized by a purity parameter.
Unlike the q-bit case where GKLS-master equation is derived by using appropriate requirements on the coupling with the environment, here we elevate the form of the master equation to a principle to describe dissipation on  the matrices of $\sl (2, \mathbb{R})$ which correspond to Gaussian kernels.
A ``pointwise'' comparison with the dynamics of Gaussian states as appears in the literature would require another paper.
Here we comment very briefly on the difference between our approach and some of the most common available in the literature.
The standard approach is to transform the ``master equation'' into a Fokker-Planck equation for the Wigner function, which means a partial differential equation. (For instance this may be found in Ref.~\cite{Ferraro-2005}.
Another approach uses the Jacobi group (a semidirect-product subgroup of the Ehrenfest group) and the associated Lie--Poisson structure (again infinite-dimensional) \cite{Bonet-Luz-2016}.

Yet, another approach is the one in Ref. \cite{Simon-1988}.
This one is the closest one to our approach, indeed  they describe the dynamics of Gaussian pure states by means of coadjoint orbits of the symplectic group on its Lie algebra.
Our approach uses the coincidence (valid only in three dimensions) that the Lie algebra $\sl(2,\mathbb{R})$ is isomorphic with $\sp(2,\mathbb{R})$ and with $\su(1,1)$.
Moreover, we describe a dynamics that goes across symplectic orbits, containing both a gradient term and a jump term.
This is completely new with respect to the approach of Simon et al., and with respect to the quoted ones, our approach describes dynamics on a finite-dimensional space by means of ordinary differential equations instead of partial differential equations.
The extent to which our approach covers all other descriptions has to be investigated further, and will be the subject of forthcoming papers.
The dynamics of pure Gaussian states may also be described in terms of second order differential geodesic equations, see Ref~\cite{Ciaglia-Fabio-2009}.


\section*{Acknowledgements}

H. Cruz-Prado is grateful for the scholarship provided by CONAHCyT M\'exico, with reference number 379177.	
O. Casta\~nos thanks support from PASPA of DGAPA-UNAM..


\section{References}
	


\end{document}